\pdfoutput=1
\documentclass[conference]{vldb}

\usepackage{graphicx}
\usepackage{balance}  
\usepackage{cite}
\usepackage{subcaption}
\usepackage{booktabs} 
\usepackage{amsmath}
\usepackage{tabularx}
\usepackage{xcite}
\usepackage{setspace}
\usepackage[linesnumbered,ruled,noend]{algorithm2e}
\usepackage{algorithmicx}
\usepackage{times}

\newcommand{\nop}[1]{}

\newtheorem{definition}{\bf Definition}
\newtheorem{problem statement}{\bf Problem Statement}

\newcommand{\iG}{\mathcal{G}}

\newcommand{\e}{\epsilon}

\vldbTitle{Dolha - an Efficient and Exact Data Structure for Streaming Graphs}
\vldbAuthors{Fan Zhang, Lei Zou, Li Zeng, Xiangyang Gou}
\vldbDOI{https:doi.org/10.14778/xxxxxxx.xxxxxxx}
\vldbVolume{xx}
\vldbNumber{xxx}
\vldbYear{2019}

\begin{document}


\title{Dolha - an Efficient and Exact Data Structure for Streaming Graphs}



%
%
%
%

\numberofauthors{4} 

\author{
%
%
\alignauthor
Fan Zhang\\
       \affaddr{Peking University}\\
       \affaddr{Beijing, China, 100080}\\
       \email{zhangfanau@pku.edu.cn}
\alignauthor
Lei Zou\\
       \affaddr{Peking University,}\\
       \affaddr{Beijing, China, 100080}\\
       \email{zoulei@pku.edu.cn}
\alignauthor 
Li Zeng\\
       \affaddr{Peking University}\\
       \affaddr{Beijing, China, 100080}\\
       \email{li.zeng@pku.edu.cn}
\and
\alignauthor 
Xiangyang Gou\\
       \affaddr{Peking University}\\
       \affaddr{Beijing, China, 100080}\\
       \email{gxy1995@pku.edu.cn}       
}

\maketitle

\begin{abstract} 
A streaming graph is a graph formed by a sequence of incoming edges with time stamps. Unlike static graphs, the streaming graph is highly dynamic and time related. In the real world, the high volume and velocity streaming graphs such as internet traffic data, social network communication data and financial transfer data are bringing challenges to the classic graph data structures. We present a new data structure: double orthogonal list in hash table (Dolha) which is a high speed and high memory efficiency graph structure applicable to streaming graph. Dolha has constant time cost for single edge and near linear space cost that we can contain billions of edges information in memory size and process an incoming edge in nanoseconds. Dolha also has linear time cost for neighborhood queries, which allow it to support most algorithms in graphs without extra cost. We also present a persistent structure based on Dolha that has the ability to handle the sliding window update and time related queries.
\end{abstract}

\section{Introduction}\label{sec:introduction}
In the real world, billions of relations and communications are created every day. A large ISP needs to deal about $10^9$  packets of network traffic data per hour per router \cite{graphsurvey2012vldb}; 100 million users log on Twitter with around 500 million tweets per day \cite{twitter}; In worldwide, the total number of sent/received emails are more than 200 billion per day \cite{email}. Those relations are coming and fading away like the tides and mining knowledge from the highly dynamic graph data is as difficult like capturing the certain wave of the sea. To handle this situation, we need a graph data structure that has high memory efficiency to contain the enormous amount of data and high speed to seize every nanosecond of the stream.

There have been several prior arts in streaming graph summarization like TCM \cite{tcm} and and specific queries like TRIÈST \cite{triest}. However, there are some complicated situations that these existing work did not cover. To illustrate our problem in this paper, we first give some motivation examples as follows:

\textbf{Use Case 1: Network traffic.} The network traffic is a typical kind of streaming graphs. Each IP address indicates one vertex and the communication between two IPs indicates an edge. Along with the data packets sending and receiving, the graph changes rapidly. To monitor this network, we need to run queries on this streaming graph. For example, to detect the suspects of cyber-attack, we want to know how many data packets each IP sends or receives and how many KBs data each edge carries. This problem is defined as vertex query and edge query and could be solved by the graph summarization system \cite{tcm} in $O(1)$ time cost. However, if we need more structure-aware query answers, such as "who are the receivers of given IP?", "who are the 2-hop neighbors of this IP?" and "how many IPs that this IP could reach?", the existing graph summarization techniques (such as TCM \cite{tcm}) cannot provide accurate query answers. In some applications, an \emph{exact} data structure is desirable for streaming graphs rather than probabilistic data structure.

\textbf{Use Case 2: Social network.} In a social network graph, a user is considered as one vertex and the relations are the edges from this user. One of the most common queries is triangle counting and there are many algorithms to deal with this problem. But existing solutions are designed specifically for triangle counting \cite{triest} and so are some continuous subgraph matching systems \cite{timingsubg} and circle detecting systems \cite{cycledetection2018vldb} over streaming graphs. If we want to run different kinds of dynamic graph analysis, we have to maintain multiple streaming systems that are costly on both space and time. An elegant solution is one uniform system that could support most graph analysis algorithms on streaming graphs.

Usually, an edge in streaming graph is received with a time-stamp indicating the edge arrive time. Some applications need to figure out \textbf{historical information} or \textbf{time constrains} based on these time stamps, however few systems support these \emph{time-related} graph queries for historical information. Here are two examples:


\textbf{Use Case 3: Financial transaction.} For example, a bank has a streaming graph system to monitor last seven days' money transactions. Each customer is recorded as one vertex, and each money tracer is recorded as one edge. On Friday, the bank receives a notice from another branch that a few suspicious transfers are made on Tuesday between 10am and 4pm from this bank.To find these suspicious transfers, the bank needs to run some pattern match on the time constrained transfers. In this case, we need a streaming graph system not only supports \emph{last snapshot}-based queries but also enable \emph{time-related} queries to figure out historical information. 

\begin{figure}[h!]
\centering
\resizebox{0.55\linewidth}{!}{
	\includegraphics{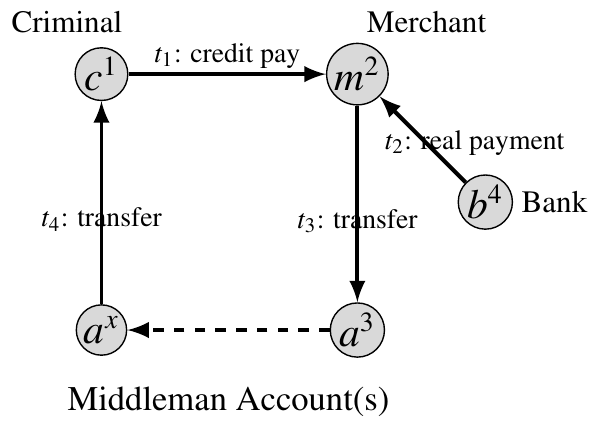}
}
\caption{Credit card  fraud in transactions (Taken from \protect \cite{cycledetection2018vldb})}
\label{fig:credit-card-fraud}
\end{figure}

\textbf{Use case 4: Fraud detection.} The same bank from Case 3 receives another report from police. The report has a list of suspicious accounts that may involves credit card fraud and the money transfer pattern they use. The bank needs to find when such pattern appeared in the transaction record among those accounts. Figure \ref{fig:credit-card-fraud} shows an example of credit card fraud pattern. In this case, we have the bank's account ID, merchant account ID and a list of suspicious accounts ID that had transactions with the merchant account. Consider these IDs as vertex and the transactions as edges, we could construct a set of query graphs. We need to locate the occurrence time when these query graphs appear in the streaming transaction graph, then we check inward and outward neighbors of these suspicious accounts near that time and find other criminal group members.

Motivated by above use cases, an efficient streaming grap structure should satisfy the following requirements: 
\begin{itemize}
	\item To enable efficient graph computing, the space cost of the data structure should be small enough to fit into main memory; 
	\item For the enormous amount of data and the high-frequency updating, the data structure must have $O(1)$ time cost to handle one incoming edge processing;
	\item The data structure should support many kinds of graph algorithms rather than designed for one specific graph algorithm;  
	\item The data structure should also support time-related queries for historical information. 
\end{itemize}


In the literature, there exist some streaming graph data structures. Generally speaking, they are classified into two categories: general streaming graph data structure and a data structure designed for some specific graph algorithms. General streaming graph structures are designed to preserve the whole structure of streaming graphs, thus, they can support most of graph algorithms like BFS, DFS, reachability query and subgraph matching by using neighbor search primitives. Most of these kind of structures are based on hash map associated with some classical graph data structures such as adjacency matrix and adjacency list. GraphStream Project \cite{graphstreamproject} is based on adjacency list associated with hash map. The basic idea of this structure is to map the vertex IDs into a hash table. Each cell of vertex hash table stores the vertex ID and the incoming/outgoing links. TCM \cite{tcm} and gMatrix \cite{gmatrix} propose to combine hash map with adjacency matrix. Different from \cite{graphstreamproject}, TCM and gMatrix are approximate data structures that inherit query errors due to hash conflicts. There are also some other streaming graph data structures that support a single \emph{specific} graph algorithm, such as HyperANF \cite{hyperanf} for t-hop neighbor and distance query, the Single-Sink DAG \cite{gao2014continuous} for pattern matching and TRIEST \cite{triest} for triangle counting. 

Table \ref{tab:general} lists the space cost of different general streaming graph data structure together with the time complexity to handle edge insertion and edge/1-hop queries. GraphStream's edge insertion time is $O(d)$, which depends on the maximum vertex degree. In many scale-free network data, the maximum vertex degree is often very large. Thus, GraphStream is not suitable for high speed streaming graph. TCM and gMatrix have the square space cost that prevents them to be used in large graphs. On the contrary, our proposed approach (called Dolha) in this paper fits all requirements for streaming graphs. Generally, Dolha is the combination of the orthogonal list with hash techniques. The orthogonal list builds two single linked lists of the outgoing and incoming edges for each vertex and store the first items of two list in vertex cell. On the other hand, the hash table is commonly used for  streaming data structure to achieve amortized $O(1)$ time look up, such as bloom filter \cite{bloom} and count-min \cite{countmin}. The combination of orthogonal list and hash table is an promising option to achieve our goal. Based on this idea, we present a new exact streaming graph structure: \emph{d}ouble \emph{o}rthogonal \emph{l}ist in \emph{ha}sh table (Dolha). 

\begin{table}[!ht]
\centering
    \caption{General Streaming Graph Structures}    
    \label{tab:general}
	\resizebox{1.01\linewidth}{!}
	{
		    \begin{large}
    \begin{tabular}{|c|c|c|c|}
    \hline
        {\bfseries } & {\bfseries Adjacency List} & {\bfseries Adjacency Matrix} & {\bfseries Orthogonal List} \\
    \hline
        {\bfseries +Hash} & {\bfseries GraphStream \cite{graphstreamproject}} & {\bfseries TCM \cite{tcm}} & {\bfseries Dolha} \\
    \hline
        Space Cost & $O(|E| \log |V|)$  & $O(|V|^2)$ & $O(|E| \log |E|)$ \\
    \hline
        Time Cost per Edge & $O(\log d)$  & $O(1)$ & $O(1)$ \\
    \hline
        Edge Query & $O(\log d)$  & $O(1)$ & $O(1)$ \\
    \hline
        1-hop Neighbor Query & $O(d)$  & $O(|V|)$ & $O(d)$ \\
    \hline
    \end{tabular}
    \end{large}

	}
\end{table}

\textbf{Our Contributions:} Table \ref{tab:general} shows the comparison among the three general streaming graph structures. In this paper:

\begin{enumerate}
	\item We design an effective data structure (Dolha) for streaming graphs with $O(|E|\log |E|)$ space cost and $O(1)$ time cost for a single edge operation. Compared with existing data structures, Dolha is more suitable in the context of high speed straming graph data. 
	\item The Dolha data structure can answer many kinds of queries over streaming graphs, among which Dolha support the query primitive edge query in $O(1)$ time and 1-hop neighbor queries in $O(d)$ time. 
	\item We present a variant of Dolha, Dolha persistent that supports sliding window and time related queries in linear time cost.
	\item Extensive experiments over both real and synthetic datasets confirm the superiority of Dolha over the state-of-the-arts. 
\end{enumerate}

\section{Related Work}
Among the existing studies, we categorize the structures into two classes: general streaming graph structures and streaming graph algorithms structures. 

\subsection{General Streaming Graph Structures}
General streaming graph structures are designed to preserve the data of graph stream and maintain the graph connection information at the same time. A general streaming graph structures could support most of graph algorithms like BFS, DFS, reachability query and subgraph matching by using neighbor search primitives. Most of these kind of structures are based on hash map associated with basic graph data structure like adjacency matrix and adjacency list. There are two different general streaming graph structures: exact structure and approximation structure. 

\textbf{Exact Structures}: Graph Stream Project \cite{graphstreamproject} is an exact graph stream processing system which is implanted by Java. Graph Stream Project is based on adjacency list associated with hash map and it supports most of graph algorithms. The basic idea of this structure is to map the vertex IDs into a hash table. Each cell of vertex hash table stores the vertex ID and the incoming / outgoing links. 

Adjacency list needs $O(|E| \log |V|)$ space and $O(|V|+|E|)$ time for traversal. However, to locate an edge, we need to go through the neighbor lists pf both in and out vertices  which indicates $O(|E|)$ time cost in some extreme situations. Even we put the neighbor list into a sorted list, it still costs $O(\log d)$ time ($d$ is the average degree of vertices) for each edge look up.

\begin{figure}[h!]
\centering
\resizebox{1.05\linewidth}{!}{
	\includegraphics{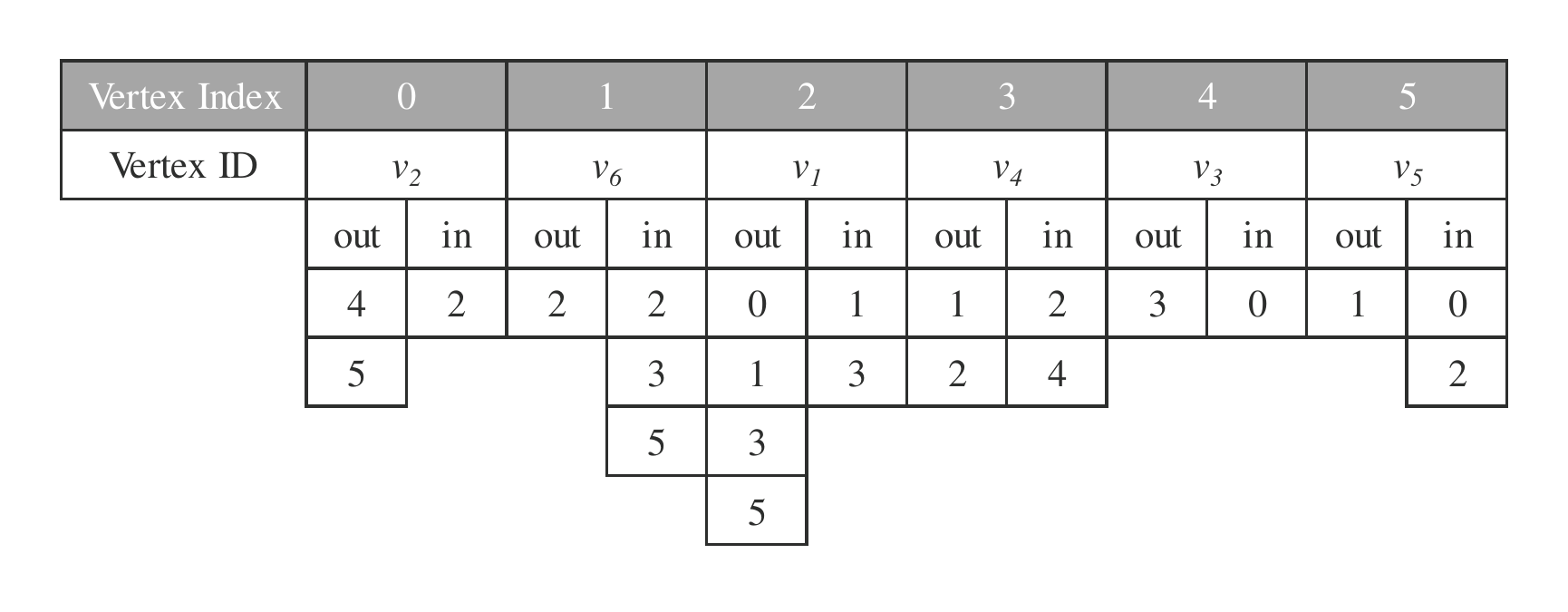}
}
\caption{Example of adjacency list in hash table}
\label{fig:adjacencylist}
\end{figure}

Figure \ref{fig:adjacencylist} shows an example of adjacency list in hash table. We use hash function $H(*)$ to map the $6$ vertices into $6$ cells vertex hash table and each cell has $2$ sorted list to store the outgoing and incoming neighbors of the vertex. i.e., $H(v_2) = 0$ and cell $0$ stores the vertex ID $v_2$, the outgoing list $\{4=H(v_3),5=H(v_5)\}$ and incoming list $\{2=H(v_1)\}$. The adjacency list stores the exact information of the graph stream but cost $O(d)$ for each edge insertion. 

\textbf{Approximation Structures}: Another solution for the structure of streaming graph is adjacency matrix in hash table. We could hash the vertices into a hash table and using a pair of vertices indexes as coordinates to construct an adjacency matrix. Vertex query in hash table is $O(1)$ time cost and so is edge look-up in the matrix. From the view of time cost, adjacency matrix in hash table is efficient but $O(|V|^2)$ space cost is a drawback. In the real world, graphs are usually sparse and we could not afford to spend $2.5$ quadrillion on a $50$ million vertices graph. There is a compromise formula that we compress the vertices into $O(\sqrt{|E|})$ size or even smaller hash table to reduce the space cost up to $O(|E|)$. But with the high compress ratio, it’s only suite for a graph summarization system, like TCM \cite{tcm}, gMatrix \cite{gmatrix}.

\begin{figure}[h!]
\centering
\resizebox{0.6\linewidth}{!}{
	\includegraphics{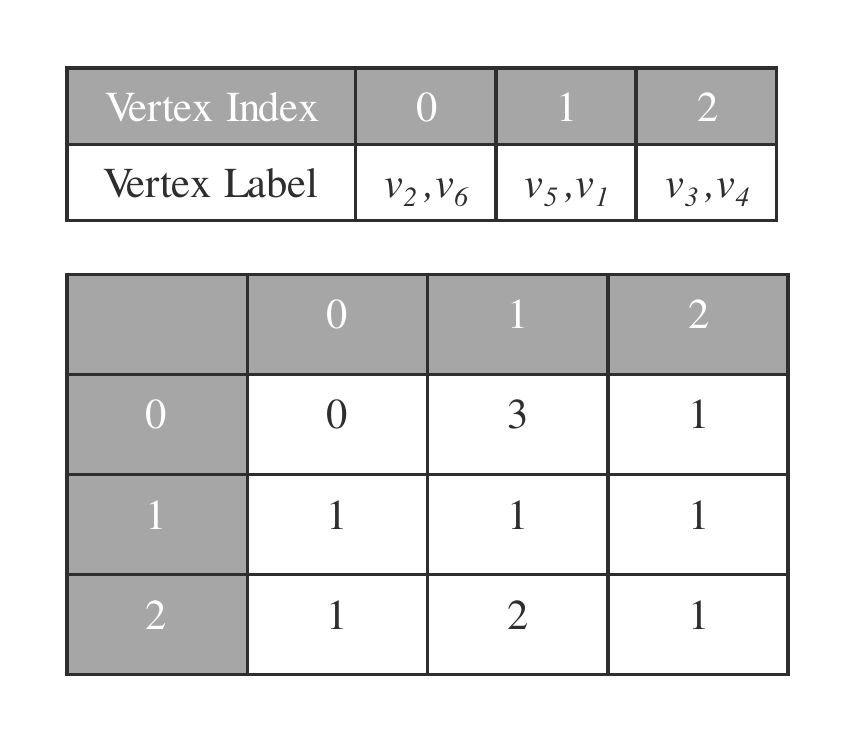}
}
\caption{Example of adjacency matrix in hash table}
\label{fig:tcm}
\end{figure}

Figure \ref{fig:tcm} shows an example of adjacency matrix in hash table. We use hash function $H(*)$ to map the $6$ vertices into $3$ cells hash table and use the table index to build a $3 \times 3$ matrix. In the $9$ cells of the matrix, we store the weights of $11$ edges. i.e., $H(v_1)=1$ and $H(v_2)=0$, the matrix table cell $(1,0)$ indicates the edge $\overrightarrow{v_1v_2}$. However, the cell $(1,0)$ also indicates the edge $\overrightarrow{v_1,v_6}$ and $\overrightarrow{v_5,v_6}$ since the hash collision. If we do outgoing neighbor query for $v_2$, the result is $\{v_5,v_1,v_2,v_3\}$ and the correct answer is $\{v_5,v_3\}$. In this case, if we want the exact result, the matrix size is $6 \times 6$ which is much larger than the edge size $11$.

\subsection{Specific Streaming Graph Structures}
Unlike the general structure, there are some data structures designed for specific algorithms on graph stream. For example, HyperANF \cite{hyperanf} is an approximation system for t-hop neighbor and distance query; the Single-Sink DAG \cite{gao2014continuous} is for pattern matching on large dynamic graph; TRIÈST \cite{triest} is sampling system for triangle counting in streaming graph; and there are some connectivity and spanners structures showed in Graph stream survey \cite{streamsurvey}. These systems could only support the designed algorithms and become incapable or unacceptable on other graph queries.

\begin{figure}[h!]
\newcommand{\mywidth}{0.35\linewidth}
\centering
	\begin{subfigure}[t]{\mywidth}
		\centering
		\resizebox{\linewidth}{!}
		{
			\includegraphics{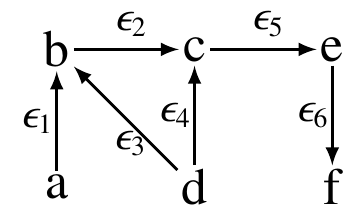}
		}
		\caption{query graph}
		\label{fig:qgraph}
	\end{subfigure}
	\begin{subfigure}[t]{\mywidth}
		\centering
		\resizebox{0.8\linewidth}{!}
		{
			\includegraphics{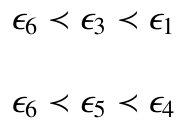}
		}
		\caption{timing order}
		\label{fig:qtiming}
	\end{subfigure}
	\caption{Running example query $Q$ (Taken from \cite{timingsubg})}
	\label{fig:examplequery}
\end{figure}

Time constrained continuous subgraph search over streaming gra- phs \cite{timingsubg} is a rare and the latest research work that considers the time as query parameter. This paper proposed an a kind of query that requires not only the structure matching but also the time order matching. Figure \ref{fig:examplequery} shows an example of time constrained subgraph query. In this query, each edge of query graph has a time-stamp constrain $\e$. A matching subgraph means the subgraph is an isomorphism of query graph and the time-stamps are following the given order.

\section{Problem Definition}\label{sec:problemdef}

\begin{definition}[Streaming Graph] \label{def:graphstream}
	A streaming graph $\iG$ is a directed graph formed by a continuous and time-evolving sequence of edges $\{\sigma_1, \sigma_2, ...\sigma_x\}$. Each edge $\sigma_i$ from vertex $u_i$ to $v_i$ is arriving at time $t_i$ with weight $w_i$, denoted as $\sigma_i(\overrightarrow {u_iv_i},t_i,w_i)$, $i=1,...,x$. 
	
\end{definition}

\begin{figure}[h!]
\centering
\resizebox{1\linewidth}{!}{
	\includegraphics{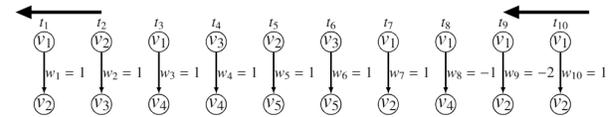}
}
\caption{Streaming Graph $S$}
\label{fig:gstream}
\end{figure}

Generally, there are two models of streaming graphs in the literature. One is only to care the latest snapshot structure, where the latest snapshot is the superposition of all coming edges to the latest time point. The other model records the historical information of the streaming graphs. The two models are formally defined in Definitions \ref{def:snapshot} and \ref{def:continuousstructure}, respectively. In this paper, we propose a uniform data structure (called \emph{Dolha}) to support both of them.  


\begin{definition}[Snapshot \& Latest Snapshot Structure] \label{def:snapshot} 

	An \\ edge $\overrightarrow {uv}$ may appear in $\iG$ multiple times with different weights at different time stamps. 
	Each occurrence of $\overrightarrow {uv}$ is denoted as $\sigma^{j}(\overrightarrow{uv},t^{j},w^{j})$, $j=1,..,n$. The total weight of edge $\overrightarrow{uv}$ at snapshot $t$ is the weight sum of all occurrences before (and including) time point $t$, 
	denoted as 
		    $$W^{t}(\overrightarrow{uv})=\sum\nolimits_{t^j\leq t} {w^j} .$$
		    where $\sigma(\overrightarrow{uv},t^{j},w^j)$ appears in streaming graph $\iG$.

	For a streaming graph $\iG$, the corresponding snapshot at time point $t$ (denoted as $\iG_t$) is a set of edges that has positive total weight at time $t$:
	    $$\iG_t=\{(\overrightarrow {uv}) \in \iG \mid W^{t}(\overrightarrow{uv})>0 \}.$$

When $t$ is the current time point, $\iG_t$ denotes the \emph{the latest snapshot structure} of $\iG$.
\end{definition}
\begin{figure}[h!]
\newcommand{\mywidth}{0.3\linewidth}
\newcommand{\mylinewidth}{\linewidth}
\centering
	\begin{subfigure}[t]{\mywidth}
		\centering
		\resizebox{\mylinewidth}{!}
		{
			\includegraphics{snapshot5}
		}
		\caption{$t = 5$}
		\label{fig:snapshot5}
	\end{subfigure}
	\hspace{0.2in}
	\begin{subfigure}[t]{\mywidth}
		\centering
		\resizebox{\mylinewidth}{!}
		{
			\includegraphics{snapshot6}
		}
		\caption{$t = 6$}
		\label{fig:snapshot6}
	\end{subfigure}
	\hspace{3in}

	\begin{subfigure}[t]{\mywidth}
		\centering
		\resizebox{\mylinewidth}{!}
		{
			\includegraphics{snapshot7}
		}
		\caption{$t = 7$}
		\label{fig:snapshot7}
	\end{subfigure}
	\hspace{0.2in}
	\begin{subfigure}[t]{\mywidth}
		\centering
		\resizebox{\mylinewidth}{!}
		{
			\includegraphics{snapshot8}
		}
		\caption{$t = 8$}
		\label{fig:snapshot8}
	\end{subfigure}
	\hspace{3in}

	\begin{subfigure}[!t]{\mywidth}
		\centering
		\resizebox{\mylinewidth}{!}
		{
			\includegraphics{snapshot9}
		}
		\caption{$t = 9$}
		\label{fig:snapshot9}
	\end{subfigure}
	\hspace{0.2in}
	\begin{subfigure}[!t]{\mywidth}
		\centering
		\resizebox{\mylinewidth}{!}
		{
			\includegraphics{snapshot10.pdf}
		}
		\caption{$t = 10$}
		\label{fig:snapshot10}
	\end{subfigure}
	\caption{ snapshot $\iG_5$ to snapshot $\iG_{10}$ of streaming graph $\iG$}
	\label{fig:snapshots}
\end{figure} 

    An example of streaming graph $\iG$ is shown in Figure \ref{fig:gstream}. Figure \ref{fig:snapshots} shows the snapshots of $\iG$ from $t_7$ to $t_{10}$. In Figure \ref{fig:snapshot7}, total edge weight $\overrightarrow{v_1v_2}$ is updated from $W^{1}(\overrightarrow{v_1v_2})=1$ (at time $t_1$) to $W^{7}(\overrightarrow{v_1v_2})=2$ (at time $t_7$). In Figure \ref{fig:snapshot8}, edge $\overrightarrow{v_1v_4}$ receives a negative weight update. Since the weight of $\overrightarrow{v_1v_4}$ is $0$ after update, it means that it is deleted from the snapshot $\iG_8$ at time $t_8$. In Figure \ref{fig:snapshot9}, the deletion of edge $\overrightarrow{v_1v_2}$ causes the deletion of vertex $v_1$ in $\iG_9$ and $v_1$ is added into $\iG_{10}$ again because the new edge $\overrightarrow{v_1v_2}$ incoming at $t_{10}$.
    

In some applications, we need to record the historical information of streaming graphs, such as fraud detection example (Use Case 4) in Section \ref{sec:introduction}. Thus, we also consider the sliding window-based model. 

\begin{definition}[Sliding Window]  \label{def:slidingWindowUpdate}
	Let $t_1$ be the starting time of a streaming graph $\iG$ and $w$ be the window length. In every update, the window would slide $\theta$ and $\theta<w$. $D^{i}_{w,\theta}(\iG)$ contains all edges in the $i$-th sliding window, denoted as:
     $$ D^{i}_{w,\theta}(\iG)=\{(\overrightarrow {uv},t,w)|$$
     $$(\overrightarrow {uv},t,w) \in \iG, t_0+(i-1) \times  \theta \le t \le t_0+(i-1) \times \theta +w\}. $$ \cite{slidingwindow}
     
\end{definition}
In Figure \ref{fig:timewindow}, the window size $w=7$ and each step the window slides $\theta=3$ edges. Figure \ref{fig:timewindow} illustrates the first and the second sliding window, where the left-most three edges expired in the second window. 

\begin{figure}[h!]
\newcommand{\mywidth}{\linewidth}
\newcommand{\mylinewidth}{\linewidth}
\centering
	
		\resizebox{\mylinewidth}{!}
		{
			\includegraphics{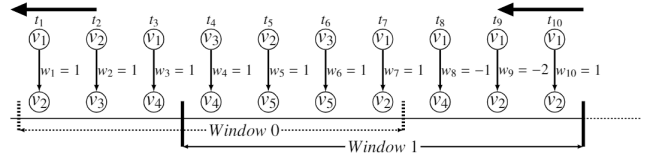}
		}
		
	\hspace{3in}
		\caption{Sliding window update on streaming graph}
		\label{fig:timewindow}
\end{figure}

\begin{definition}[Window Based Persistent Structure]  \label{def:continuousstructure}
Given a streaming graph $\iG$, the \textbf{Window Based Persistent Structure} (``persistent structure'' for short) is a graph formed by all the unexpired edges in the current time window. Each edge is associated with the time stamps denoting the arriving times of the edge. An edge may have multiple time stamps due to the multiple occurrences.

\end{definition}
\begin{figure}[h!]
\centering
\resizebox{0.5\linewidth}{!}{
	\includegraphics{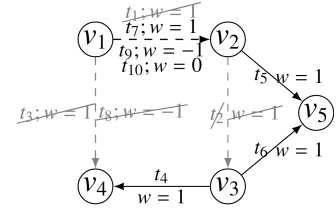}
}
\caption{Window based persistent structure}
\label{fig:timeupdate}
\end{figure}
In a snapshot streaming graph structure, only the latest snapshot is recorded and the historical information is overwritten. For example, a snapshot structure only stores the snapshot ${\iG}_{10}$ at last time point $t_{10}$ in Figure \ref{fig:snapshot10}. The update process of the streaming graph is overwritten. 

Assume that the second time window (Window 1) is the current window. Figure \ref{fig:timeupdate} shows how the persistent structure stores the streaming graph. Edge $\overrightarrow {v_1v_2}$ is associated with three time points ($t_7$, $t_9$ and $t_{10}$) that are all in the current time window. Although edge $\overrightarrow {v_1v_2}$ also occurs at time $t_1$, it is expired in this time window. The gray edges denotes all expired edges, such as $\overrightarrow {v_1v_4}$ and $\overrightarrow {v_2v_3}$.




\begin{definition}[Streaming graph query primitives]
\label{def:query}
	We define 4 query primitives for streaming graph $\iG$ and most of the graph algorithms such as DFS, BFS, reachability query and subgraph matching are based on these query primitives:
\begin{enumerate}
	\item \textbf{Edge Query:}
	Given the a pair of vertices IDs $(u,v)$, return the weight or time stamp of the edge $\overrightarrow {uv}$. If the edge doesn’t exist, return $\{null\}$.
	\item \textbf{Vertex Query:}
	Given the a vertex IDs $u$, return the incoming or outgoing weight of $u$. If the vertex does not exist, return $\{null\}$.
	\item \textbf{1-hop Successor Query:}
	Given the a vertex IDs $u$, return a set of vertices that $u$ could reach in 1-hop. If there is no such vertex, return $\{null\}$.
	\item \textbf{1-hop Precursor Query:}
	Given the a vertex IDs $u$, return a set of vertices that could reach $u$ in 1-hop. If there is no such vertex, return $\{null\}$.
\end{enumerate}
\end{definition}
    The query primitives are slightly different in two structures. If we query edge $\overrightarrow {v_1v_2}$ in snapshot structure at $\iG_{10}$, the result is the last updated edge information : $(\overrightarrow {v_1v_2},t_{10},1)$. If we query edge $\overrightarrow {v_1v_2}$ in persistent structure at $\iG_{10}$ showing in Figure \ref{fig:timeupdate}, the result is a list of unexpired edges:
    {$(\overrightarrow {v_1v_2},t_7,0)$, $(\overrightarrow {v_1v_2},t_9,-1)$, $(\overrightarrow {v_1v_2},t_{10},1)$}. The same difference applies to 1-hop successor query and precursor query. If we query the successor of $v_1$ at $t_{10}$, the snapshot structure will give the answer $v_2$. But the persistent structure will return a set of answers: {$(v_2,t_7)$, $(v_2,t_9)$, $(v_2,t_{10})$}. 
  
    Based on the persistent structure query primitives, we define a new type of queries on streaming graph named \emph{time related query} that considers the time stamps as query parameters. In this paper, we adopt two kinds of time related queries: time constrained pattern query is to find the match subgraph in a given time period; structure constrained time query is to find the time periods that given subgraph appears in $\iG$.
  
\begin{definition}[Time Constrained Pattern Query]\label{def:subgraphmatch} 
	A pattern \\ graph is a triple $P=(V(P), E(P), L)$, where $V(P)$ is a set of vertices in $P$, $E(P)$ is a set of directed edges, $L$ is a function that assigns a label for each vertex in $V (P)$. Given a pattern graph $P$ and a time period $(t,t')$ and $t<t'$, $\iG$ is a time constrained pattern match of $P$ if and only if there exists a bijective function $F$ from $V(P)$ to $V(g)$ such that the following conditions hold:
\begin{enumerate}
	\item \textbf{Structure Constraint (Isomorphism)}
		\begin{itemize}
		\item
			$\forall u \in V(P),L(u) = L(F(u))$.
		\item
			$\overrightarrow {uv}  \in E(P) \Leftrightarrow \overrightarrow {F(u)F(v)}  \in E(g)$.
		\end{itemize}
	\item   \textbf{Time Period Constraint} 
	\begin{itemize}
		\item
			$\forall \overrightarrow{uv} \in E(P),t \le  t_{\overrightarrow{uv}}  \le t'.$\cite{timingsubg}
	\end{itemize}			
\end{enumerate}
    In this paper, the problem is to find all the time constrained pattern matches of given $P$ over $\iG_{t'}$ which is the snapshot of $\iG$ at time $t'$. 
\end{definition}

\begin{figure}[h!]
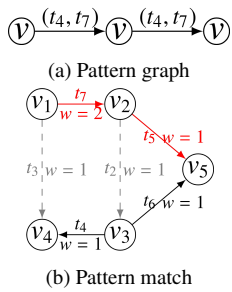

\newcommand{\mywidth}{\linewidth}
\newcommand{\mylinewidth}{\linewidth}
\centering
	\begin{subfigure}[t]{\mywidth}
		\centering
		\resizebox{0.35\mylinewidth}{!}
		{
			\includegraphics{pattern}
		}
		\caption{Pattern graph}
		\label{fig:pattern}
	\end{subfigure}
	\hspace{3in}
	\begin{subfigure}[t]{\mywidth}
		\centering
		\resizebox{0.3\mylinewidth}{!}
		{
			\includegraphics{patternmatch}
		}
		\caption{Pattern match}
		\label{fig:patternmatch}
	\end{subfigure}
	\caption{Time constrained pattern query}
	\label{fig:patternquery}
\end{figure} 
    Figure \ref{fig:patternquery} shows an example of time constrained pattern query. In Figure \ref{fig:pattern}, a pattern graph is given which queries all the 2-hop connected structures. The edges of pattern graph have a time constrain that only the edges with the time stamp between $(t_4,t_7)$ are considered as match candidates. Figure \ref{fig:patternmatch} is the snapshot $\iG_7$ of $\iG$ at time $t_7$. Edge $\overrightarrow{v_1v_4}$ and $\overrightarrow{v_2v_3}$ are discarded since the time stamps are out of time constrain. Edge set $\{(\overrightarrow{v_1v_2})(\overrightarrow{v_2v_5})\}$ is the only matching subgraph for the given pattern on $\iG$.
    
\begin{definition}[Structure Constrained Time Quer\\y]\label{def:timematch} 
	A query graph $Q$ is a sequence of directed edges $\{q_1,q_2,...,q_m\}$ and $T$ is a set of time pairs $\{(t_1,t^{'}_{1}),...,(t_n,t^{'}_{n})\}$. Given a pattern graph $Q$, a structure constrained time match $T$ is that $Q$ is the subgraph of every snapshot of $\iG$ during any time period $(t_i,t^{'}_{i})$ in $T$.
		$$\forall t, t_i \le t \le t^{'}_{i}, Q \in G_t. $$
	
\end{definition}
  
    \begin{figure}[h!]
\centering
\resizebox{0.15\linewidth}{!}{
	\includegraphics{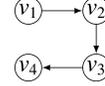}
}
\caption{Structure constrained time query}
\label{fig:timequery}
\end{figure}
    Figure \ref{fig:timequery} gives an example of structure constrained time query edge set $\{\overrightarrow{v_1v_2},\overrightarrow{v_2v_3},\overrightarrow{v_3v_4}\}$ is given. On $\iG$, the query graph is the subgraph of every snapshot from $\iG_4$ to $\iG_8$ until deletion of $\overrightarrow{v_1v_2}$ on $\iG_9$. In $\iG_{10}$, the query graph is matching again since the new arriving $\overrightarrow{v_1v_2}$. The query result of Figure \ref{fig:timequery} is $\{(t_4,t_7),(t_{10},t_{10})\}$.

\section{DOLHA - DOUBLE ORTHOGONAL LIST IN HASH TABLE} \label{sec:baseline}
    \begin{table}[!ht]
    \centering
    \small
    \caption{Notations}     
    \label{tab:notations}
	\resizebox{0.85\linewidth}{!}
	{
		    \begin{small}
    \begin{tabular}{|l|l|l|l|}
    \hline
        {\bfseries Notation} & \multicolumn{1} {c|} {\bfseries Definition and Description} \\
   \hline
        $G_s$ / $G_t$        & Streaming graph / Snapshot at time point $t$\\
   \hline
         $D_s$ / $D_p$     &  Dolha snapshot / Dolha persisdent\\
    \hline
        $\overrightarrow{uv}$  & The directed  edge from vertex $u$ to $v$\\
   \hline
        Doll        & Doulble orthogonal linked list \\
   \hline
        $O$         & Outgoing Doll \\
   \hline
        $I$         & Incoming Doll \\
    \hline
        $T$         & Time travel linked list \\
   \hline
        $w$         & Edge weight\\
    \hline
        $t$         & Edge time stamp\\
    \hline
        $H(*)$      & Hash value of $*$\\
    \hline
        $V(*)$      & Vertex table index of $*$\\
    \hline
        $E(*)$      & Edge table index of $*$\\
    \hline
        $E^{*}_A()$           & First item's edge table index of link $*$\\
    \hline
        $E^{*}_{\Omega}()$    & Last item's edge table index of link $*$\\
    \hline
        $E^{*}_{\Omega}()$    & Last item's edge table index of link $*$\\
    \hline
        $E^{*}_N()$    & Next item's edge table index of link $*$\\
    \hline
        $E^{*}_P()$    & Previous item's edge table index of link $*$\\
    \hline
        $*^{-/+}$    & Previous/next item of $*$\\
    \hline
    \end{tabular}
    \end{small}

	}
\end{table}

\begin{figure}[h!]
    \centering
    \resizebox{0.8\linewidth}{!}{
	\includegraphics{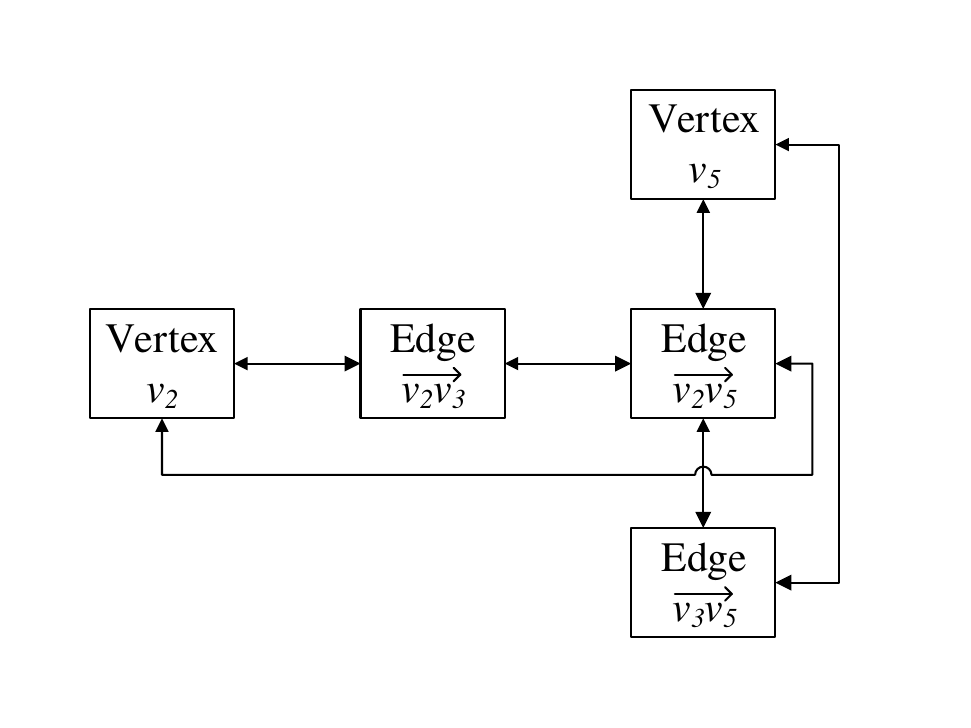}
    }
    \caption{Example of Doll}
    \label{fig:doll}
    \end{figure}

In order to handle high speed streaming graph data, we propose the data structure---called Double Orthogonal List in Hash Table (\emph{Dolha} for short)---in this paper. Essentially, Dolha is the combination of double orthogonal linked list with hash tables. A \emph{d}ouble \emph{o}rthogonal \emph{l}inked \emph{l}ist (\emph{Doll} for short) is a classical data structure to store a graph, in which each edge $\overrightarrow{uv}$ in graph $\iG$ is both in the double linked list of all the outgoing edges from vertex $u$: $\{\overrightarrow{uv_A},...{\overrightarrow{uv_\Omega}}\}$ denotes as \emph{outgoing Doll} and in the double linked list of all the incoming edges to vertex $v$: $\{\overrightarrow{u_Av},...{\overrightarrow{u_\Omega v}}\}$ denotes as \emph{incoming Doll}. Vertex $u$ has two pointers to the first item $v_A$ and last item $v_\Omega$ of outgoing Doll. Vertex $v$ has two pointers to the first item $u_A$ and last item $u_\Omega$ of incoming Doll. For example, Figure \ref{fig:doll} illustrates an example of Doll.

\nop{
creates a double orthogonal linked list for each vertex and edge, denoted as Doll, and map the vertices and edges into hash tables. By using this structure, we can construct an exact snapshot structure $D_s$ for streaming graph $\iG$ depending. 
    
    \textbf{Doll - double orthogonal linked list} Doll is a data structure to store a graph. For any edge $\overrightarrow{uv}$ in Doll of graph $\iG$, $\overrightarrow{uv}$ is both in the double linked list of all the outgoing edges from vertex $u$: $\{\overrightarrow{uv_A},...{\overrightarrow{uv_\Omega}}\}$ denotes as outgoing Doll and in the double linked list of all the incoming edges to vertex $v$: $\{\overrightarrow{u_Av},...{\overrightarrow{u_\Omega v}}\}$ denotes as incoming Doll. Vertex $u$ has two pointers to the first item $v_A$ and last item $v_\Omega$ of outgoing Doll. Vertex $v$ has two pointers to the first item $u_A$ and last item $u_\Omega$ of incoming Doll. 

}    

\nop{
As shown in Table \ref{tab:notations}, we denote function $H(*)$ to get the hash value for an item, function $V(u)$ to get vertex table index of $u$ and $E(\overrightarrow {uv})$ to get edge table index of $\overrightarrow {uv}$. The superscript of function indicates the linked list which the item belongs to. There are four linked links that an item could be: hash collision linked list $H$, outgoing Doll $O$, incoming Doll $I$ and time travel linked list $T$. The subscript of function indicates the item's location on the linked list: the next item $N$, the previous item $P$, the first item $A$, the last item $\Omega$. The plus and minus on the item indicate the next or previous item on certain link. 
    
For example, $E^{O}_{N}(\overrightarrow {uv}^{-})$ indicates the next item's edge index of $\overrightarrow {uv}$'s previous item on outgoing Doll which is the edge index of $\overrightarrow {uv}$: $$E^{O}_{N}(\overrightarrow {uv}^{-})==E(\overrightarrow {uv})$$

We construct $D_s$ and $D_s$ is the latest snapshot of $\iG$. $D_s$ costs $O(ElogE)$ space and the insertion, deletion or update operation on $D_s$ cost constant time. On $D_s$, we can perform all 4 query primitives and get exact result. The edge query and vertex query cost constant time; 1-hop Successor Query and 1-hop Precursor Query cost $log|d|$ time.
}

\subsection{Dolha Snapshot Data Structure}
Given a graph $\iG$, the Dolha structure contains of four key-value tables. Before that, we assume that each vertex $u$ (and edge $\overrightarrow{uv}$) is hashed to a hash value $H(u)$ (and $H(\overrightarrow{uv})$). For example, we use hash function $H(*)$ to map the vertices and edges:
 \begin{itemize}
		\item 
		$H(v_1)=1, H(v_2)=2,H(v_3)=0, H(v_4)=1, H(v_5)=3$
		\item
		$H(\overrightarrow{v_1v_2})=1, H(\overrightarrow{v_2v_3})=0, H(\overrightarrow{v_1v_4})=4, H(\overrightarrow{v_3v_4})=2, H(\overrightarrow{v_2v_5})=4, H(\overrightarrow{v_3v_5})=3$
\end{itemize}

\nop{
Tables \ref{tab:vertexmap} and \ref{tab:edgemap} show the hash values of each vertex and edge. 
\begin{table}[!ht]
\centering
\small
    \caption{Hash map for Vertices}     
    \label{tab:vertexmap}
	\resizebox{0.7\linewidth}{!}
	{
		    \begin{small}
    \begin{tabular}{|c|c|c|c|c|c|}
    \hline
        {\bfseries  } & {\bfseries $H(v_1)$} & {\bfseries $H(v_2)$} & {\bfseries $H(v_3)$} & {\bfseries $H(v_4)$} & {\bfseries $H(v_5)$} \\
    \hline
        = & $1$ & $2$ & $0$ & $1$ & $3$ \\
    \hline
    \end{tabular}
    \end{small}

	}
\end{table}

\begin{table}[!ht]
\centering
\small
    \caption{Hash map for edges}    
    \label{tab:edgemap}
	\resizebox{1\linewidth}{!}
	{
		    \begin{normalsize}
    \begin{tabular}{|c|c|c|c|c|c|c|}
    \hline
        {\bfseries} & {\bfseries \small $H(\overrightarrow{v_1v_2})$} & {\bfseries \small $H(\overrightarrow{v_2v_3})$} & {\bfseries \small $H(\overrightarrow{v_1v_4})$} & {\bfseries $H(\overrightarrow{v_3v_4})$} & {\bfseries \small $H(\overrightarrow{v_2v_5})$} & {\bfseries \small  $H(\overrightarrow{v_3v_5})$}\\
    \hline
        = & $1$ & $0$ & $4$ & $2$ & $4$ & $3$\\
    \hline
    \end{tabular}
    \end{normalsize}

	}
\end{table}
}

\textbf{Vertex Hash Table:} Dolha creates $m_v(m_v \ge |V|)$ size vertex hash table and uses function $H(*)$ map the vertex ID $u$ to vertex hash table index $H(u)$. Due to the hash collision, there could be a list of vertices with same hash table index. In each table cell, Dolha stores the vertex table index of the first vertex on collision list.

Table \ref{tab:vertexhash} is an example of vertex hash table. We use $H(v_1)=1$ as hash index to locate the vertex table index $0$ and find the $v_1$'s details in vertex table cell $0$. The vertex $v_4$ has the same hash value as $v_1$ which means the hash collision occurs. We use hash value $1$ to find the first vertex $v_1$ on the collision list then we can find the next item $v_4$'s vertex table index $3$ in $v_1$'s vertex table cell.

\textbf{Vertex Table $V$:} Dolha creates $m_v(m_v \ge |V|)$ size vertex table and one empty cell variable denoted as $\phi_V$. Initially, $\phi_V=0$ . We denote the vertex table index for new coming vertex $u$ as $V(u)$. Let $V(u)=\phi_V$ and increase $\phi_V$ by $1$. In each vertex table cell, Dolha stores the vertex ID, the outgoing weight sum $w_O(u)$ and incoming weight sum $w_I (u)$, the head and tail edge table index for outgoing Doll \nop {$E^{O}_{A}(u)$ and $E^{O}_{\Omega}(u)$}, the head and tail edge table index for incoming Doll \nop {$E^{I}_{A}(u)$ and $E^{I}_{\Omega}(u)$} and the vertex table index of the next vertex on collision list \nop {$V^{H}_{N}(u)$}.

Table \ref{tab:vertextable} shows the vertex table of $\iG_5$ in Figure \ref{fig:snapshots}. Out/In $w$ indicates the outgoing and incoming weights of the vertex. $O$ is the edge table index of first and last items of outgoing Doll and $I$ is the edge table index of first and last items of incoming Doll. $H$ is the next vertex on the collision list. The vertices are given indexes incrementally ordered by first arriving time. $\phi_V=5$ means vertex table is full. If more vertices arrive, we can create a new vertex table and begin with index $5$ as the extension of existing vertex table. 

\textbf{Edge Hash Table:} Edge hash table: Dolha creates $m_e(m_e \ge |E|)$ size vertex hash table and uses function $H(*)$ map the outgoing vertex ID $u$ plus incoming vertex ID $v$ of edge $\overrightarrow {uv}$ to edge hash table index $H(\overrightarrow {uv})$. Same as the vertex hash table, Dolha stores the edge table index of the first edge on collision list \nop {$E^{H}_{A}(\overrightarrow {uv})$}.

In Table \ref{tab:edgehash}, we have the same method as vertex hash table to deal with hash collision. $\overrightarrow{v_1v_4}$ has the same hash value $4$ as $\overrightarrow{v_2v_5}$. In cell $4$, we can find $\overrightarrow{v_1v_4}$'s edge table index $2$ then find $\overrightarrow{v_2v_5}$'s edge table index.

\textbf{Edge Table $E$}: Dolha creates $m_e(m_e \ge |E|)$ size vertex table and one empty cell flag denoted as $\phi_E$. Initially, $\phi_E=0$. We denote the vertex table index for new coming edge $\overrightarrow {uv}$ as $E(\overrightarrow {uv})$. Let $E(\overrightarrow {uv})=\phi_E$ and increase $\phi_E$ by 1. In each edge table cell, Dolha stores the vertex table indexes $V(u)$ and $V(v)$, the weight $w(\overrightarrow {uv})$, the time stamp $t(\overrightarrow {uv})$, the previous and next edge table index for outgoing Doll \nop {$E^{O}_{P}(u)$ and $E^{O}_{N}(u)$}, the previous and next edge table index for incoming Doll \nop {$E^{I}_{P}(u)$ and $E^{I}_{N}(u)$} and the edge table index of the next edge on collision list \nop {$E^{H}_{N}(\overrightarrow {uv})$}.

Table \ref{tab:edgetable} shows the edge table of $\iG_5$ in Figure \ref{fig:snapshots}.
$w$ is the weight and $t$ is the time stamp. Vertex index indicates the outgoing and incoming vertices of the edge. $O$ is the edge table index of next and previous items of outgoing Doll and $I$ is the edge table index of next and previous items of incoming Doll. $H$ is the next edge on the collision list.   

\begin{table}[!ht]
\centering
\small
    \caption{Vertex hash table of $\iG_5$}     
    \label{tab:vertexhash}
	\resizebox{0.7\linewidth}{!}
	{
		    \begin{small}
    \begin{tabular}{|c|c|c|c|c|c|}
    \hline
        {\bfseries Hash index} & {\bfseries 0} & {\bfseries 1} & {\bfseries 2} & {\bfseries 3} & {\bfseries 4} \\
    \hline
        Vertex table index & $2$ & $0$ & $1$ & $4$ & $/$ \\
    \hline
    \end{tabular}
    \end{small}

	}
\end{table}

\begin{table}[!ht]
\centering
\small
    \caption{Vertex table of $\iG_5$}    
    \label{tab:vertextable}
	\resizebox{0.9\linewidth}{!}
	{
		    \begin{small}
    \begin{tabular}{|c|c|c|c|c|c|c|c|c|c|c|}
    \hline
         {\bfseries Index} & \multicolumn{2}{|c|}{\bfseries 0} & \multicolumn{2}{|c|}{\bfseries 1} & \multicolumn{2}{|c|}{\bfseries 2} & \multicolumn{2}{|c|}{\bfseries 3} & \multicolumn{2}{|c|}{\bfseries 4} \\
    \hline
        Vertex ID & \multicolumn{2}{|c|}{$v_1$} & \multicolumn{2}{|c|}{$v_2$} & \multicolumn{2}{|c|}{$v_3$} & \multicolumn{2}{|c|}{$v_4$} & \multicolumn{2}{|c|}{$v_5$} \\
    
     \hline
        Out/In $w$ & $2$ & $0$   & $2$ & $1$   & $1$ & $1$   & $0$ & $2$&   $0$ & $1$    \\
         
    \hline
        O & $0$ & $2$   & $1$ & $4$   & $3$ & $3$   & $/$ & $/$&   $/$ & $/$    \\
    \hline
        I & $/$ & $/$   & $0$ & $0$   & $1$ & $1$   & $2$ & $3$&   $4$ & $4$    \\
    \hline
        H & \multicolumn{2}{|c|}{$3$} & \multicolumn{2}{|c|}{$/$} & \multicolumn{2}{|c|}{$/$} & \multicolumn{2}{|c|}{$/$} & \multicolumn{2}{|c|}{$/$} \\
    \hline
        
    \end{tabular}
    \end{small}
	}
	$\phi_V=5$
\end{table}

\begin{table}[!ht]
\centering
\small
    \caption{Edge hash table of $\iG_5$}     
    \label{tab:edgehash}
	\resizebox{0.7\linewidth}{!}
	{
		    \begin{small}
    \begin{tabular}{|c|c|c|c|c|c|c|}
    \hline
        {\bfseries Hash index} & {\bfseries 0} & {\bfseries 1} & {\bfseries 2} & {\bfseries 3} & {\bfseries 4} & {\bfseries 5}\\
    \hline
        Edge table index & $1$ & $0$ & $3$ & $/$ & $2$ & $/$\\
    \hline
    \end{tabular}
    \end{small}

	}
\end{table}

\begin{table}[!ht]
\centering
\small
    \caption{Edge table of $\iG_5$}     
    \label{tab:edgetable}
	\resizebox{1.05\linewidth}{!}
	{

    \begin{small}
    \begin{tabular}{|c|c|c|c|c|c|c|c|c|c|c|c|c|}
    \hline
         {\bfseries Index} & \multicolumn{2}{|c|}{\bfseries 0} & \multicolumn{2}{|c|}{\bfseries 1} & \multicolumn{2}{|c|}{\bfseries 2} & \multicolumn{2}{|c|}{\bfseries 3} & \multicolumn{2}{|c|}{\bfseries 4} & \multicolumn{2}{|c|}{\bfseries 5}\\
         
     \hline
         {\bfseries $w$} & \multicolumn{2}{|c|}{\bfseries 1} & \multicolumn{2}{|c|}{\bfseries 1} & \multicolumn{2}{|c|}{\bfseries 1} & \multicolumn{2}{|c|}{\bfseries 1} & \multicolumn{2}{|c|}{\bfseries 1} & \multicolumn{2}{|c|}{\bfseries /}\\
         
      \hline
         {\bfseries $t$} &  \multicolumn{2}{|c|}{\bfseries 1} & \multicolumn{2}{|c|}{\bfseries 2} & \multicolumn{2}{|c|}{\bfseries 3} & \multicolumn{2}{|c|}{\bfseries 4} & \multicolumn{2}{|c|}{\bfseries 5} &\multicolumn{2}{|c|}{\bfseries /} \\
         
    \hline
        Vertex index & $0$ & $1$ & $1$ & $2$ & $0$ & $3$ & $2$ & $3$& $1$ & $4$& $/$ & $/$\\
    \hline
        O & $/$ & $2$   & $/$ & $4$   & $0$ & $/$   & $/$ & $/$&   $1$ & $/$ &    $/$ & $/$\\
    \hline
        I & $/$ & $/$   & $/$ & $/$   & $/$ & $3$   & $2$ & $/$&   $/$ & $/$ &    $/$ & $/$\\
    
    \hline 
        H & \multicolumn{2}{|c|}{$/$} & \multicolumn{2}{|c|}{$/$} & \multicolumn{2}{|c|}{$4$} & \multicolumn{2}{|c|}{$/$} & \multicolumn{2}{|c|}{$/$} & \multicolumn{2}{|c|}{$/$} \\
    \hline    
    \end{tabular}
    \end{small}
	}
	$\phi_E=5$
\end{table}
\nop{
    Table \ref{tab:vertexhash}, \ref{tab:vertextable}, \ref{tab:edgehash} and \ref{tab:edgetable} show the Dolha snapshot of $\iG_6$. In table \ref{tab:vertexmap}, vertex IDs are mapped into hash values by hash function $H(*)$, so are the edges in table \ref{tab:edgemap}. The hash values indicate the hash indexes in hash tables \ref{tab:vertexhash} and \ref{tab:edgehash}. The vertex table index in vertex hash table points to the vertex table index of first item hashed into that cell and so does the edge hash table. For example,  $v_1$ is hashed into vertex hash table cell $1$ and we can find the vertex table index $V(1)$ of $v_1$. $v_4$ is also hashed into cell $1$. We go to vertex table cell $1$ and find it is not $v_4$. Then we use the hash collision linked list and find index $V(3)$ which is the location of $v_4$. The edge $\overrightarrow{v_1v_4}$ and $\overrightarrow{v_2v_5}$ use the same mechanism to deal with the hash collision.
    
    In Table \ref{tab:vertextable}, $O$ is the edge table index of first and last items of outgoing Doll and $I$ is the edge table index of first and last items of incoming Doll. In Table \ref{tab:edgetable}, $O$ is the edge table index of next and previous items of outgoing Doll and $I$ is the edge table index of next and previous items of incoming Doll. 
}
\subsection{Dolha Snapshot Construction}
    When an edge $(\overrightarrow {uv};t;w)$ comes:
    \begin{itemize}
		\item
			Map the edge $\overrightarrow {uv}$ into edge hash table cell $H(\overrightarrow {uv})$.
		\item
		    If $H(\overrightarrow {uv})$ is empty, $\overrightarrow {uv}$ does not exist in $D_s$. If $H(\overrightarrow {uv})$ is not empty, traverse the collision list of cell $H(\overrightarrow {uv})$ in edge hash table. If find $\overrightarrow {uv}$, $\overrightarrow {uv}$ exists; if not, $\overrightarrow {uv}$ does not exist.
	\end{itemize}
	
	There are two possible operations:
	
	\textbf{If $\overrightarrow {uv}$ does not exist in $D_s$:}
	\begin{itemize}
		\item
		   Add $\overrightarrow {uv}$ into into edge table cell $E(\overrightarrow {uv})$ and the collision list of $H(\overrightarrow {uv})$.
		\item
			Map the vertices $u$,$v$ into vertex hash table $H(u)$,$H(v)$.
		\item
		    If $H(u)$ is empty, add ID $u$ into vertex table cell $V(u)$. If $H(u)$ is not empty, traverse the collision list of cell $H(u)$ in vertex hash table. If find match ID, then we update vertex table $V(u)$ of $u$; if not, add $u$ into vertex table cell $V(u)$ and collision list of $H(u)$. 
		 \item
		    Do the same operation for $v$.
		 \item
            Add $\overrightarrow {uv}$ into the end of outgoing Doll of $u$ and incoming Doll of $v$.
	\end{itemize}
	
	\textbf{If $\overrightarrow {uv}$ exists in $D_s$:}
	\begin{itemize}
		\item
			Set $t(\overrightarrow {uv})=t$ and $w(\overrightarrow {uv})=w(\overrightarrow {uv})+w$.
			
		\item
			Delete $\overrightarrow {uv}$ from outgoing Doll of $u$ and incoming Doll of $v$ 
		\item
			If $\overrightarrow {uv}$ has positive weight after this update:
		\item
			Add $\overrightarrow {uv}$ into the end of outgoing and incoming Dolls.
		\item
	    	if $\overrightarrow {uv}$ has zero or negative weight after this update:
		\item
			Delete $\overrightarrow {uv}$ from edge table.
		\item
            If there is not any item in both Doll of $u$ or $v$, delete $u$ or $v$.
	\end{itemize}

\begin{algorithm}[h!]
\small
\caption{Dolha snapshot edge processing}
    \label{alg:snapshot}
\KwIn{Streaming graph $\iG$}
\KwOut{Dolha snapshot structure of $\iG$}
\For{each incoming edge $(\overrightarrow {uv};t;w)$ of $\iG$}
{
    \textbf{Check existence of $\overrightarrow {uv}$:} \\
    Map $\overrightarrow {uv}$ into $H(\overrightarrow {uv})$. \\
	\If{$H(\overrightarrow {uv})$ is $null$}
	{
	    $\overrightarrow {uv}$ does not exist\\
	} 
	\Else
    {
        Traverse the collision list from $E^H_{A} (\overrightarrow {uv})$. \\
        \If {reach $null$ and no match for $\overrightarrow {uv}$}
        {
         $\overrightarrow {uv}$ does not exist\\
        }
        \Else
        {
        $\overrightarrow {uv}$ exists\\
        }
    }
    \If{$\overrightarrow {uv}$ does not exist}
    {
        \textbf{Update collision list of $\overrightarrow {uv}$:} \\
        \If{$H(\overrightarrow {uv})$ is empty}
        {
            Let $E^{H}_{A}(\overrightarrow {uv})=E(\overrightarrow {uv})$\\
        }
        \Else 
        {
            Let $E^H_{N} (\overrightarrow {uv}^{-})=E(\overrightarrow {uv})$\\
        }
        \textbf{Check existence of $u$:} \\
        Map the vertices $u$ into $H(u)$\\
        \If{$H(u)$ is $null$}
        {
            Add $u$ into vertex table $V(u)$ and
            let $V^{H}_{A}(u)=V(u)$\\
        }
        \Else
        {
            Traverse the collision list from $E^H_{A} (u)$. \\
            \If {reach $null$ and no match for $u$}
            {
                Add $u$ into vertex table $V(u)$ and 
                let $V^{H}_{N}(u^{-})=V(u)$\\
            }
        }
        \textbf{Do the same operation for $v$ same as $u$} \\
         Add $\overrightarrow {uv}$ into edge table $E(\overrightarrow {uv})$\\
         \textbf{Add $\overrightarrow {uv}$ into outgoing Doll:} \\
        \If{both $E^{O}_{A}(u)$ and $E^{O}_{\Omega}(u)$ are $null$}
        {
            Let $E^{O}_{A}(u)=E(\overrightarrow {uv}$) and $E^{O}_{\Omega}(u)=E(\overrightarrow {uv})$\\
        }
        \If{neither $E^{O}_{A}(u)$ nor $E^{O}_{\Omega}(u)$ is $null$}
        {
            Let $E^{O}(\overrightarrow {uv}^{-})=E^{O}_{\Omega}(u)$ and 
             $E^{O}_{N}(\overrightarrow {uv}^{-})=E(\overrightarrow {uv})$ and 
             $E^{O}_{P}(\overrightarrow {uv})=E^{O}(\overrightarrow {uv}^{-})$ and 
             $E^{O}_{\Omega}(u)=E(\overrightarrow {uv})$\\
        }
        \textbf{Add $\overrightarrow {uv}$ into incoming Doll same as outgoing Doll} \\
        
    }
    \If{$\overrightarrow {uv}$ exists}
    {
        Let $w(\overrightarrow {uv})+=w$ and $t(\overrightarrow {uv})=t$ \\
        \textbf{Delete $\overrightarrow{uv}$ from outgoing Doll:} \\
        \If{$\overrightarrow {uv}$ is the first item of outgoing Doll}
        {
            Let $E^{O}_{A}(\overrightarrow {uv})=E^{O}_{N}(\overrightarrow {uv})$ and 
            $E^{O}_{P}(\overrightarrow {uv}^{+})=null$ \\
        }
        \If{$\overrightarrow {uv}$ is the last item of outgoing Doll}
        {
            Let $E^{O}_{\Omega}(\overrightarrow {uv})=E^{O}_{P}(\overrightarrow {uv})$ and 
            $E^{O}_{N}(\overrightarrow {uv}^{-})=null$ \\
        }
        \Else
        {
            Let $E^{O}_{N}(\overrightarrow {uv}^{-})=E^{O}_{N}(\overrightarrow {uv})$ and $E^{O}_{P}(\overrightarrow {uv}^{+})=E^{O}_{P}(\overrightarrow {uv})$\\
        }
        \textbf{Delete $\overrightarrow {uv}$ from incoming Doll same as outgoing Doll} \\
        \If{$w(\overrightarrow {uv})>0$}
        {
            Add $E(\overrightarrow {uv})$ into the end of outgoing Doll and incoming Doll \\
        }
        \Else
        {
            \textbf{Delete $\overrightarrow {uv}$} \\
            Delete $E(\overrightarrow {uv})$ and flag $E(\overrightarrow {uv})$ as empty cell\\
            \If{there is no item on outgoing Doll or incoming Doll of $u$}
            {
                Delete $V(u)$ and flag $V(u)$ as empty cell\\
            }
            \If{there is no item on outgoing Doll or incoming Doll of $v$}
            {
                Delete $V(v)$ and flag $V(v)$ as empty cell\\
            }
        }
    }
}	
\end{algorithm}
    For example, at time $6$, edge $\overrightarrow{v_3v_5}$ is received. We use $H(v_3v_5)=3$ to get the edge hash table index and find edge $\overrightarrow{v_3v_5}$ is a new edge. We write the empty cell index $5$ of edge table into hash table and check the two vertices by using vertex hash table. We locate the $V(2)$ for $v_3$ and $V(4)$ for $v_5$ on vertex table and get the last item of outgoing Doll $E(3)$ and the last item of incoming Doll $E(4)$. We update the both last items of outgoing and incoming Doll to $5$ then move to edge table. We update the next item of outgoing Doll to $5$ in $E(3)$ and update the next item of incoming Doll to $5$ in $E(4)$. Finally, we write $w$, $t$, $(2,4)$, $(3,/)$ and $(4,/)$ into $E(5)$. 
    
    At time $7$, edge $\overrightarrow{v_1v_2}$ comes and it is already on the edge table. We first update the $w$ and $t$ at $E(0)$ and remove $\overrightarrow{v_1v_2}$ from both of the Dolls then add it to the end of Dolls.
    
    At time $8$, edge $\overrightarrow{v_1v_4}$ carries negative weight and $w$ is $0$ after the update. We move $E(2)$ from the outgoing and incoming doll and update the associated indexes, then we empty the cell $2$ of edge table and put the index $2$ into empty edge cell list. At time $9$, edge $\overrightarrow{v_1v_2}$ is deleted and $v_1$ has no out or in edges. We empty cell $0$ of vertex table and put the index $0$ into empty vertex cell list.
    
\subsection{Time and Space Cost}

\subsubsection{Time Cost}\label{sec:baseline:tcq}
Algorithm \ref{alg:snapshot} shows how Dolha process one incoming edge. 

From line $3$ to $14$, we maintain the edge hash table to check the existence of incoming edge $\overrightarrow {uv}$. According to \cite{hashcollision}, if we hash n items into a hash table of size n, the expected maximum list length is $O(\log n / \log\log n)$. In the experiment, more than $99\%$ collision list is less than $\log n / \log\log n$, more than $90\%$ collision list is shorter than $5$. Hash table could achieve amortized $O(1)$ time cost for $1$ item insert, delete and update which is much faster than sorted table. This step costs $O(1)$ time.

If $\overrightarrow {uv}$ is a new edge, from line $16$ to $22$, we maintain the vertex hash table to check the existence of two vertices $u$ and $v$. In this step, we do two hash table look up and it costs $O(1)$ time. From line $23$ to $29$, we write $\overrightarrow {uv}$ into edge table then add it into the end of outgoing and incoming Dolls. The time complexity of this step is same as insertion on double linked list which is $O(1)$.

If $\overrightarrow {uv}$ exists, from line $31$ to $38$, we update the weight and time stamp of $\overrightarrow {uv}$ then delete it from outgoing and incoming Dolls. This step costs the same time as deletion on double linked list which is also $O(1)$. From line $39$ to $40$, if updated weight is positive, we add the $\overrightarrow {uv}$ to the end of both two Dolls which costs $O(1)$. If the updated weight is zero or negative, we delete $\overrightarrow {uv}$ completely then delete $u$ and $v$ if they have $0$ in and out degrees. Line $41$ to $46$ shows the deletions and this step also costs $O(1)$.

Overall, for each incoming edge processing, the time complexity of Dolha is $O(1)$.

\subsubsection{Space Cost}\label{sec:baseline:tcq}
    Dolha snapshot structure needs one $|V|$ cells vertex hash table, one $|V|$ cells vertex table, one $|E|$ cells edge hash table and one $|E|$ cells edge table. Dolha also needs a $\log |V|$ bits integer for one vertex index and $\log |E|$ bits for one edge index. 
    
    \textbf{Vertex hash table:} Each cell only stores one vertex index. It costs $\log |V| \times |V|$ space.
    
    \textbf{Edge hash table:} Each cell only stores one edge index. It costs $\log |E| \times |E|$ space.
    
    \textbf{Vertex table:} Each cell stores vertex ID, in and out weights one $\log |V|$ bits vertex index for collision list, four $\log |E|$ bits edge indexes for Dolls. It costs $(\log |V| + 4  \times  \log |E|) \times |V|$ space.
    
    \textbf{Edge table:} Each cell stores weight, time stamp, one $\log |E|$ bits edge index for collision list, two $\log |V|$ bits vertex index for in and out vertices, four $\log |E|$ bits edge indexes for Dolls. It costs $(2 \times \log |V| + 5  \times  \log |E|) \times |E|$ space.
    
    In total, Dolha needs $(2 \times \log |V|+4 \times \log |E|) \times |V|+(2 \times \log |V|+5 \times \log |E|) \times |E|$ bits for the data structure. Since usually $|V| \ll |E|$, the space cost of Dolha snapshot structure is $O(|E|\log |E|)$. 
\section{Dolha Persistent Structure} \label{sec:baseline}
\subsection{Dolha Persistent Data Structure}
Using Dolha, We could construct a persistent structure $D_p$ and $D_p$ contains any snapshot's information of $\iG$. $D_p$ has the same structure as $D_s$ except the time travel list.

\begin{definition}[Time Travel List]  \label{def:timetravellist}
	An edge $\overrightarrow {uv}$ may appear in streaming graph $S$ multiple times with different time stamp. Time travel list $T$ is a single linked list that links all the edges $\overrightarrow {uv}$ which share same outgoing and incoming vertices. In $T$, each edge has an index points to its previous appearance in the stream.
\end{definition}

$D_p$ also has four index-value tables. The vertex hash table, vertex table and edge hash table are same as $D_s$. In each cell of edge table, $D_p$ has a extra value \nop {$E^T_{P}(*)$} which indicates the previous item on the time travel list.

\subsection{Dolha Persistent Construction}
\subsubsection{Incoming Edge Processing}
When an edge $\sigma(\overrightarrow {uv};t;w)$ comes: 
\begin{itemize}
		\item
            Check the existence of $\overrightarrow {uv}$ same as Dolha snapshot.
	\end{itemize}

	 \textbf{If $\overrightarrow {uv}$ does not exist in $D_p$:}
	\begin{itemize}
		\item
            The operation is exact same as Dolha snapshot.
	\end{itemize}
		
	 \textbf{If $\overrightarrow {uv}$ exists in $D_p$:}
	\begin{itemize}
	    \item
	    Use edge hash table to find the existing edge table index $E(\sigma ')$ of $\overrightarrow {uv}$.
		\item
		    Insert edge $\sigma$ as new edge into edge table and set the time travel list index as $E(\sigma ')$.
		\item
		    Update the edge table index of $\overrightarrow {uv}$ on edge hash collision list. 
	\end{itemize}

\begin{algorithm}[h!]
\small
\caption{Dolha persistent edge processing}
    \label{alg:persistent}
\KwIn{Streaming graph $\iG$}
\KwOut{Dolha persistent structure of $\iG$}
\For{each incoming edge $\sigma(\overrightarrow {uv};t;w)$ of $\iG$}
{
    \textbf{Check existence of $\overrightarrow {uv}$:} \\
    \If{$\overrightarrow {uv}$ does not exist}
    {
        Insert $\overrightarrow {uv}$\\
    }
    \If{$\overrightarrow {uv}$ exists in cell $E(\sigma ')$}
    {
       Insert $E(\sigma)$ as new edge and let $w(\sigma)=w(\sigma)+w(\sigma ')$ \\
       Let $E^T_{P}(\sigma)=E(\sigma ')$\\
    }
    \If{value of $H(\overrightarrow {uv})$ in edge hash table is $null$}
    {
        Let $E^H_A(\overrightarrow {uv})=E(\sigma)$\\
    }
    \Else
    {
        Let $E^H_N(\overrightarrow {uv}^{-})=E(\sigma)$
    }
}	

\end{algorithm}

\nop{
\begin{table}[!ht]
\centering
\small
    \caption{Vertex table of Window 0}    
    \label{tab:vertextime}
	\resizebox{0.8\linewidth}{!}
	{
		    \begin{small}
    \begin{tabular}{|c|c|c|c|c|c|c|c|c|c|c|}
    \hline
         {\bfseries Index} & \multicolumn{2}{|c|}{\bfseries 0} & \multicolumn{2}{|c|}{\bfseries 1} & \multicolumn{2}{|c|}{\bfseries 2} & \multicolumn{2}{|c|}{\bfseries 3} & \multicolumn{2}{|c|}{\bfseries 4} \\
    \hline
        Vertex ID & \multicolumn{2}{|c|}{$v_1$} & \multicolumn{2}{|c|}{$v_2$} & \multicolumn{2}{|c|}{$v_3$} & \multicolumn{2}{|c|}{$v_4$} & \multicolumn{2}{|c|}{$v_5$} \\
        
    \hline
        O & $0$ & $6$   & $1$ & $4$   & $3$ & $5$   & $/$ & $/$&   $/$ & $/$    \\
    \hline
        I & $/$ & $/$   & $0$ & $6$   & $1$ & $1$   & $2$ & $3$&   $4$ & $5$    \\
    \hline
        H & \multicolumn{2}{|c|}{$3$} & \multicolumn{2}{|c|}{$/$} & \multicolumn{2}{|c|}{$/$} & \multicolumn{2}{|c|}{$/$} & \multicolumn{2}{|c|}{$/$} \\
    \hline
        
    \end{tabular}
    \end{small}
	}
\end{table}
}
\begin{table}[!ht]
\centering
\small
    \caption{Edge hash table of Window 0}    
    \label{tab:edgetimehash}
	\resizebox{1\linewidth}{!}
	{
		    \begin{small}
    \begin{tabular}{|c|c|c|c|c|c|c|c|c|c|c|}
    \hline
        {\bfseries Hash index} & {\bfseries 0} & {\bfseries 1} & {\bfseries 2} & {\bfseries 3} & {\bfseries 4} & {\bfseries 5} & {\bfseries 6} & {\bfseries 7} & {\bfseries 8} & {\bfseries 9}\\
    \hline
        Edge table Index & $1$ & $/$ & $3$ & $/$ & $5$ & $4$ & $/$ & $2$ & $6$ & $/$\\
    \hline
    \end{tabular}
    \end{small}

	}
\end{table}

\begin{table}[!ht]
\centering
\large
    \caption{Edge table of Window 0}     
    \label{tab:edgetime}
	\resizebox{1.05\linewidth}{!}
	{
		    \begin{small}
    \begin{tabular}{|c|c|c|c|c|c|c|c|c|c|c|c|c|c|c|c|c|c|c|c|c|}
    \hline
         {\bfseries Index} & \multicolumn{2}{|c|}{\bfseries 0} & \multicolumn{2}{|c|}{\bfseries 1} & \multicolumn{2}{|c|}{\bfseries 2} & \multicolumn{2}{|c|}{\bfseries 3} & \multicolumn{2}{|c|}{\bfseries 4} & \multicolumn{2}{|c|}{\bfseries 5} & \multicolumn{2}{|c|}{\bfseries 6} & \multicolumn{2}{|c|}{\bfseries 7} & \multicolumn{2}{|c|}{\bfseries 8} & \multicolumn{2}{|c|}{\bfseries 9}\\
         
     \hline
         {\bfseries $w$} & \multicolumn{2}{|c|}{\bfseries 1} & \multicolumn{2}{|c|}{\bfseries 1} & \multicolumn{2}{|c|}{\bfseries 1} & \multicolumn{2}{|c|}{\bfseries 1} & \multicolumn{2}{|c|}{\bfseries 1} & \multicolumn{2}{|c|}{\bfseries 1} 
         & \multicolumn{2}{|c|}{\bfseries 2} & \multicolumn{2}{|c|}{\bfseries /} & \multicolumn{2}{|c|}{\bfseries /} & \multicolumn{2}{|c|}{\bfseries /}\\
         
      \hline
         {\bfseries $t$} &  \multicolumn{2}{|c|}{\bfseries 1} & \multicolumn{2}{|c|}{\bfseries 2} & \multicolumn{2}{|c|}{\bfseries 3} & \multicolumn{2}{|c|}{\bfseries 4} & \multicolumn{2}{|c|}{\bfseries 5} &\multicolumn{2}{|c|}{\bfseries 6} & \multicolumn{2}{|c|}{\bfseries 7} & \multicolumn{2}{|c|}{\bfseries /} & \multicolumn{2}{|c|}{\bfseries /} & \multicolumn{2}{|c|}{\bfseries /}\\
         
    \hline
        V & $0$ & $1$  &  $1$ & $2$ & $0$ & $3$ & $2$ & $3$& $1$ & $4$& $2$ & $4$ & $0$ & $1$ & $/$ & $/$ & $/$ & $/$ & $/$ & $/$\\
    \hline
        O & $/$ & $2$   & $/$ & $4$   & $0$ & $6$   & $/$ & $5$&   $1$ & $/$ &    $3$ & $/$  & $2$ & $7$  & $/$ & $/$ & $/$ & $/$ & $/$ & $/$\\
    \hline
        I & $/$ & $/$   & $/$ & $6$   & $/$ & $3$   & $2$ & $7$&   $/$ & $5$ &    $4$ & $/$ & $1$ & $8$ & $/$ & $/$ & $/$ & $/$ & $/$ & $/$\\
    \hline
        H & \multicolumn{2}{|c|}{$/$} & \multicolumn{2}{|c|}{$/$} & \multicolumn{2}{|c|}{$/$} & \multicolumn{2}{|c|}{$/$} & \multicolumn{2}{|c|}{$/$} & \multicolumn{2}{|c|}{$/$} & \multicolumn{2}{|c|}{$/$} & \multicolumn{2}{|c|}{$/$} & \multicolumn{2}{|c|}{$/$} & \multicolumn{2}{|c|}{$/$}\\
        
    \hline
        T & \multicolumn{2}{|c|}{$/$} & \multicolumn{2}{|c|}{$/$} & \multicolumn{2}{|c|}{$/$} & \multicolumn{2}{|c|}{$/$} & \multicolumn{2}{|c|}{$/$} & \multicolumn{2}{|c|}{$/$} & \multicolumn{2}{|c|}{$0$} & \multicolumn{2}{|c|}{$/$} & \multicolumn{2}{|c|}{$/$} & \multicolumn{2}{|c|}{$/$}\\

    \hline    
    \end{tabular}
    \end{small}
	}
\end{table}

\begin{table}[!ht]
\centering
\large
    \caption{Edge table of Window 1}     
    \label{tab:edgewin1}
	\resizebox{1.05\linewidth}{!}
	{
		    \begin{small}
    \begin{tabular}{|c|c|c|c|c|c|c|c|c|c|c|c|c|c|c|c|c|c|c|c|c|}
    \hline
         {\bfseries Index} & \multicolumn{2}{|c|}{\bfseries 0} & \multicolumn{2}{|c|}{\bfseries 1} & \multicolumn{2}{|c|}{\bfseries 2} & \multicolumn{2}{|c|}{\bfseries 3} & \multicolumn{2}{|c|}{\bfseries 4} & \multicolumn{2}{|c|}{\bfseries 5} & \multicolumn{2}{|c|}{\bfseries 6} & \multicolumn{2}{|c|}{\bfseries 7} & \multicolumn{2}{|c|}{\bfseries 8} & \multicolumn{2}{|c|}{\bfseries 9}\\
         
     \hline
         {\bfseries $w$} & \multicolumn{2}{|c|}{\bfseries /} & \multicolumn{2}{|c|}{\bfseries /} & \multicolumn{2}{|c|}{\bfseries /} & \multicolumn{2}{|c|}{\bfseries 1} & \multicolumn{2}{|c|}{\bfseries 1} & \multicolumn{2}{|c|}{\bfseries 1} 
         & \multicolumn{2}{|c|}{\bfseries 1} & \multicolumn{2}{|c|}{\bfseries -1} & \multicolumn{2}{|c|}{\bfseries 0} & \multicolumn{2}{|c|}{\bfseries /}\\
         
      \hline
         {\bfseries $t$} &  \multicolumn{2}{|c|}{\bfseries /} & \multicolumn{2}{|c|}{\bfseries /} & \multicolumn{2}{|c|}{\bfseries /} & \multicolumn{2}{|c|}{\bfseries 4} & \multicolumn{2}{|c|}{\bfseries 5} &\multicolumn{2}{|c|}{\bfseries 6} & \multicolumn{2}{|c|}{\bfseries 7} & \multicolumn{2}{|c|}{\bfseries 9} & \multicolumn{2}{|c|}{\bfseries 10} & \multicolumn{2}{|c|}{\bfseries /}\\
         
    \hline
        V & $/$ & $/$  &  $/$ & $/$ & $/$ & $/$ & $2$ & $3$& $1$ & $4$& $2$ & $4$ & $0$ & $1$ & $0$ & $1$ & $0$ & $1$ & $/$ & $/$\\
    \hline
        O & $/$ & $/$  &  $/$ & $/$ & $/$ & $/$   & $/$ & $5$&   $/$ & $/$ &    $3$ & $/$  & $/$ & $7$  & $6$ & $8$  & $7$ & $/$  & $/$ & $/$\\
    \hline
        I & $/$ & $/$  &  $/$ & $/$ & $/$ & $/$   & $/$ & $/$&   $/$ & $5$ &    $4$ & $/$ & $/$ & $7$ & $6$ & $8$ & $7$ & $/$ & $/$ & $/$ \\
    \hline
        H & \multicolumn{2}{|c|}{$/$} & \multicolumn{2}{|c|}{$/$} & \multicolumn{2}{|c|}{$/$} & \multicolumn{2}{|c|}{$/$} & \multicolumn{2}{|c|}{$/$} & \multicolumn{2}{|c|}{$/$} & \multicolumn{2}{|c|}{$/$} & \multicolumn{2}{|c|}{$/$} & \multicolumn{2}{|c|}{$/$} & \multicolumn{2}{|c|}{$/$}\\
        
    \hline
        T & \multicolumn{2}{|c|}{$/$} & \multicolumn{2}{|c|}{$/$} & \multicolumn{2}{|c|}{$/$} & \multicolumn{2}{|c|}{$/$} & \multicolumn{2}{|c|}{$/$} & \multicolumn{2}{|c|}{$/$} & \multicolumn{2}{|c|}{$/$} & \multicolumn{2}{|c|}{$6$} & \multicolumn{2}{|c|}{$7$} & \multicolumn{2}{|c|}{$/$}\\

    \hline    
    \end{tabular}
    \end{small}
	}
\end{table}

Table \ref{tab:edgetimehash} and \ref{tab:edgetime} show the Dolha persistent's edge hash table and edge table of $\iG$ in Window 0. The vertex hash table and vertex table of  Dolha persistent are similar like Dolha snapshot and so is the new edge coming. But for edge $\overrightarrow{v_1v_2}$ update at time $6$, we add the update as new edge into $E(6)$ and update the edge hash table to latest update. By using the time travel list, all the updates of $\overrightarrow{v_1v_2}$ are linked. 

\subsubsection{Sliding Window Update}
When the window slides the $i$th step, we have the start time $t_s=t_0+(i-2) \times \theta$ and end time $t_e=t_0+(i-1) \times \theta$ of expired edges which need to delete from edge table. Since the edge table is naturally ordered by time, we can find the last expired edge denote as $E(\sigma_e)$ at $t_e$ in $O(\log S)$ time. By using edge hash table, we can find the latest update of $E(\sigma_\Omega)$ and traversal back by the time travel list. For each $E(\sigma_n) (e < n \le \Omega)$ on time travel list, let $w_n=w_n-w_e$. If each $w_n \le 0$, delete all the $E(\sigma_n)$. Then delete each $E(\sigma_m) (0 < m \le e)$ on time travel list. Do the same operation for the edges from $t_e$ to $t_s$. For every deleted edge, if it is the first or last item of Doll, update the associated cell in vertex table and set the index to $null$. If all the Doll indexes are $null$ in that vertex cell, delete the vertex and flag the cell as empty.

As shown in Figure \ref{fig:timewindow}, when window slides from $0$ to $1$ means the edges before $t_4$ will expire. First, we can binary search the edge table to locate the first unexpired edge index $3$ since the table is sorted by time stamp. Then we start to delete the expired edges from cell $3$. We use the hash table to check if there are unexpired updates for the expired edges. For example, $\overrightarrow{v_1v_2}$ has unexpired update at time $7$, so we minus the expired weight at cell $6$. 

Table \ref{tab:edgewin1} shows the edge table of Dolha persistent at Window 1. The first 3 expired edges have been deleted. At time $8$, $\overrightarrow{v_1v_4}$ with negative weight arrives, but there is no positive $\overrightarrow{v_1v_4}$ in this window. In this case, we won't save $\overrightarrow{v_1v_4}$. At time $9$ and $10$, $\overrightarrow{v_1v_2}$ has negative or zero weights, but $\overrightarrow{v_1v_4}$ has positive weight at time $7$, so we keep the record and link them with time travel linked list. 

\textbf{Space Recycle:} Due to the chronological ordered edge table, the expired edges are always continuous and in the head of the unexpired edges. We could always recycle the space from expired edges which means we won't need infinite space to save the continuous streaming but only need the maximum number of edges in each window. For instance, in table \ref{tab:edgewin1}, we can re-use the cell from $0$ to $1$ for next window update and we have enough space as long as no more than $9$ edges in $1$ window.

\subsection{Time and Space Cost}
The time cost of Dolha persistent is hash table cost, Doll cost and time travel list cost. For each incoming edge, the hash table cost and Doll cost are $O(1)$ as we discussed in Dolha snapshot and the time travel list cost is also $O(1)$ same as insertion on single linked list. Overall, the time cost for one edge processing is $O(1)$.

 To store all the information of streaming $S$, Dolha persistent structure needs one $|V|$ cells vertex hash table, one $|V|$ cells vertex table, one $|S|$ cells edge hash table and one $|S|$ cells edge table. In total, Dolha needs $(2 \times \log |V|+4 \times \log |S|) \times |V|+(2 \times \log |V|+5 \times \log |S|) \times |S|$ bits plus $\log |S| \times |S|$ for time travel list. The space cost of Dolha persistent structure is $O(|S|\log |S|)$.


\section{Algorithms on Dolha} \label{sec:compression}

In this section, we discuss how to perform the graph algorithms on both Dolha snapshot structure and persistent structure.

\subsection{Algorithms on Dolha Snapshot} \label{sec:mstree:definition}

\subsubsection{Query Primitives}
Dolha snapshot structure supports all the 4 graph query primitives.

\textbf{Edge Query:}
Given a pair of vertices IDs $(u,v)$, to query the weight and time stamp of edge $(\overrightarrow{uv})$ is same as the existence checking of $(\overrightarrow{uv})$ in insertion operation. By using edge hash table, we can find $E(\overrightarrow{uv})$ on edge table and return $w$ and $t$. As we proved before, the time cost of hash table checking is amortized $O(1)$.

\textbf{Vertex Query:}
Similar as edge query, by using vertex hash table, we can locate given vertex $u$ on vertex table in $O(1)$ time and return the query result.

\textbf{1-hop Successor Query and 1-hop Precursor Query:}
Given a vertex ID $u$, Dolha first perform vertex query to find $V(u)$ in $O(1)$ time. Then we have the head edge index $E^O_{A}(u)$ of outgoing Doll. From $E(\sigma)=E^O_{A}(u)$, we can use $E^O_{N}(\sigma)$ to acquire all edges on outgoing Doll iteratively and add the incoming vertex indexes of these edges into set $\{V(v)\}$. The IDs of $\{V(v)\}$ can be found in vertex table and returned as the results of 1-hop successor query. The 1-hop precursor query is similar as successor query but use incoming Doll instead. The time cost of Doll iteration depends on the outgoing or incoming degree $d$ of given $u$. The total time cost of 1-hop successor query or 1-hop precursor query is $O(d)$. 

\textbf{Chronological Doll:} 
In Dolha structure, we maintain the Doll in chronological order. The result list of 1-hop successor query or 1-hop precursor query is sorted by the time stamps. The chronological Doll could reduce the search space in some time related queries. For example, in Figure \ref{fig:examplequery}, we have a candidate edge $(\overrightarrow{uv};t)$ that matches $(\overrightarrow{dc};\e_4)$ and look for the candidate edges of $(\overrightarrow{ce};\e_5)$. Since the timing order constrain $\e_5 \prec \e_4$, we first check the time stamp of first edge on $v$'s outgoing Doll in $O(1)$ time. If the time stamp is equal or larger than $t$, it means there is no match for $(\overrightarrow{ce};\e_5)$.  If the time stamp is less than $t$, we can search from the first edge on $v$'s outgoing doll until equal or larger the time stamp than $t$.

\subsubsection{Directed Triangle Finding}

By using the 4 graph query primitives, most graph algorithms could run on Dolha. The 1-hop successor query and 1-hop precursor query associated with edge query could support all the BFS or DFS based algorithms like reachability query, tree parsing, shortest path query, subgraph matching and triangle finding. For example, the triangle finding is a common graph query on streaming graph.

To query the directed triangle on Dolha, we can use the edge iterator method. During the Dolha snapshot construction, we can add one out degree counter and one in degree counter for each vertex. For each edge $(\overrightarrow{uv})$ incoming edge, get the minimal candidate set $\{j\}$ between $v$'s successor set and $u$'s precursor set. Then check each $j$ in set $\{j\}$ that if there is $(\overrightarrow{ju})$ or $(\overrightarrow{vj})$ existing in edge table by using edge query. The set of all existing $(\overrightarrow{uv}, \overrightarrow{vj}, \overrightarrow{ju})$ is the query result.According to \cite{triangle}, the time complexity of triangle finding on whole graph is $O(\sum_{\overrightarrow{uv} \in E } \min \{d_{in}(u),d_{out}(v)\})$, so the time cost is $O(\min \{d_{in}(u),d_{out}(v)\})$ for each edge update.

\begin{algorithm}[h!]
\small
\caption{Continuous directed triangle finding on Dolha snapshot}
    \label{alg:triangle}
\KwIn{Dolha snapshot structure of $\iG$ with out and in degree counter}
\KwIn{Streaming Graph $\iG$}
\KwOut{Directed triangles in $\iG$}
\For{each new coming edge $\overrightarrow{uv}$ of $\iG$} 
{
	\If{in degree of $u$ $\le$ out degree of $v$}
	{
    	\For{each vertex $j$ in $u$'s precursor set}
	    {
	       \If{$\overrightarrow{vj}$ exsits in edge table}
	       {
	            Put $(\overrightarrow{uv},\overrightarrow{ju},\overrightarrow{vj})$ into result set\\
	       }
	    }
	}
	\Else
	{
	\For{each vertex $j$ in $v$'s successor set }
	    {
	       \If{$\overrightarrow{ju}$ exsits in edge table}
	       {
	            Put $(\overrightarrow{uv},\overrightarrow{ju},\overrightarrow{vj})$ into result set \\
	       }
	    }
	}
}
\end{algorithm}

\subsection{Algorithms on Dolha Persistent} \label{sec:mstree:definition}

\subsubsection{Query Primitives}
Dolha persistent structure also supports all the 4 graph query primitives both on the latest snapshot and persistent perspective of $\iG$:
	
\textbf{Edge Query:}
Given a pair of vertices IDs $(u,v)$, the latest update of edge $\overrightarrow{uv}$ could be found by using edge hash table. Once find the latest update of edge $\overrightarrow{uv}$, we could use time travel list to retrieve all the updates of $\overrightarrow{uv}$ in current window. 

\textbf{Vertex Query:}
The vertex query on Dolha persistent is exactly same as snapshot structure.

\textbf{1-hop Successor Query and 1-hop Precursor Query:}
Given a vertex ID $u$, the outgoing or incoming Doll of $u$ may contain duplicates of edges. To query the successor of $u$ on Dolha persistent, it's better from the last item of outgoing Doll $E^O_{\Omega}(u)$ which is definitely the latest outgoing edge from $u$. Let $E(\overrightarrow{uv})=E^O_{\Omega}(u)$, we add $v$ to the result set and use the time travel link of $\overrightarrow{uv}$ to flag all the previous update records of $\overrightarrow{uv}$. Then we traversal the outgoing doll and do the same operation for each unflagged edge as $\overrightarrow{uv}$. 1-hop precursor query is same as successor query but using the incoming Doll. The two lists are sorted by time naturally. 

\subsubsection{Time Related Queries}
\textbf{Time Constrained Pattern Query:}
Given time period $(t,t')$, the essential part of time constrained pattern query is to find the all the edges with time stamp $(t \le t_{\overrightarrow{uv}} \le t')$ on snapshot $G_t'$. The chronological edge table allows us to locate the first edge $E(\sigma_t)$ at time $t$ and the last edge $E(\sigma_t')$ at time $t'$ in $O(\log S)$ time. Then we can run Algorithm  \ref{alg:time1} to construct the adjacency list of the candidate subgraph of time constrained pattern query. We also could construct a Dolha snapshot structure to store the candidate subgraph by using Algorithm \ref{alg:time2}. The time cost of candidate subgraph construction is $O(\log S+S')$ and the space cost is $O(S')$ ($S'$ is the incoming edge number of $(t,t')$). We can run any isomorphism algorithm on the candidate subgraph structure to get the final query result.

\begin{algorithm}[h!]
\small
\caption{Adjacency list construction for candidate subgraph of time constrained pattern query}
    \label{alg:time1}
\KwIn{edges between $\sigma_t$ and $\sigma_t'$ in edge table}
\KwIn{Dolha persistent structure of $\iG$}
\KwOut{Adjacency list of candidate subgraph}
\For{each edge $\sigma(\overrightarrow{uv};t;w)$ from $E(\sigma_t')$ to $E(\sigma_t)$} 
{
	\If{$flag!=3$}
	{
    	\For{each edge on the time travel list after $\sigma$}
	    {
	        Let $flag=3$ \\
	    }
	   \If{$flag==2$ or $flag==0$}
	   {
	        Put $u$ into candidate vertex set \\
	        \For{each edge $\sigma_O$ on outgoing Doll of $u$}
	       {
	            \If{$flag!=3$}
	            {
	            	  \For{each edge on the time travel list after $\sigma_O$}
	                   {
	                       Let $flag=3$ \\
	                   }
	            	       Let $flag+=1$ \\
	            	       Put the incoming vertex of $\sigma_O$ into the outgoing neighbor list of $u$\\
	            }
	       }
	   }
	   \If{$flag==1$ or $flag==0$}
	   {
	        Put $v$ into candidate vertex set \\
	        \For{each edge $\sigma_I$ on incoming Doll of $v$}
	       {
	            \If{$flag!=3$}
	            {
	            	  \For{each edge on the time travel list after $\sigma_I$}
	                   {
	                       Let $flag=3$ \\
	                   }
	            	       Let $flag+=2$ \\
	            	       Put the outgoing vertex of $\sigma_I$ into the incoming neighbor list of $v$\\
	            }
	       }
	   }
	}
}
\end{algorithm}

\begin{algorithm}[h!]
\small
\caption{Dolha snapshot construction for candidate subgraph of time constrained pattern query}
    \label{alg:time2}
\KwIn{edges between $\sigma_t$ and $\sigma_t'$ in edge table}
\KwIn{Dolha persistent structure of $\iG$}
\KwOut{Dolha snapshot of candidate subgraph}
Construct a Dolha snapshot structure $D_t'$ with vertex and edge size $|E(\sigma_t')-E(\sigma_t)|$
\For{each edge $\sigma(\overrightarrow{uv};t;w)$ from $E(\sigma_t')$ to $E(\sigma_t)$} 
{
	\If{$flag!=1$}
	{ 
    	\For{each edge on the time travel list after $\sigma$}
	    {
	        Let $flag=1$ \\
	    }
	    Insert $\sigma$ into $D_t'$ \\
	}
}
\end{algorithm}

\textbf{Structure Constrained Time Query:}
Given a sequence of directed edges $Q\{q_1,q_2,...,q_m\}$, for each edge $q_n$ in $Q$, we can use edge hash table to locate the latest update $E(q_n)$ in $\iG$ and use time travel list to find the time period set $T_n$ that edge $q_n$ appears. Then we join all the time period sets to find result time period set. The Algorithm \ref{alg:time3} shows that the time complexity is $O(m \times p \times  \log (m \times p))$ ($p$ is the average number of one edge appearance in $S$).

\begin{algorithm}[h!]
\small
\caption{Structure constrained time query}
    \label{alg:time3}
\KwIn{a sequence of directed query edges $Q\{q_1,q_2,...,q_m\}$}
\KwIn{Dolha persistent structure of $\iG$}
\KwOut{Time period set $T$ that match the query structure}
Let $t_e=null$ and $t_s=null$\\
Let chronological order set $T_c=\phi$\\
\For{each edge $q_n$ in $Q$} 
{
    Use edge hash table to find the latest update $E(q_n)$\\
    \For{each edge $q^t_n$ on time travel list of $q_n$ from $E(q_n)$} 
    {
        \If{$w^t_n>0$ }
        {
            Let $t_s=t^t_n$\\
            \If{$t_e==null$}
            {
                 Let $t_e=t^t_n$\\
            }
        }
        \If{$w^t_n \le 0$ }
        {
            \If{$t_s!=null$}
            {
                 Put $t_s$ flag as $s$ and $t_e$ flag as $e$ into $T_c$\\
                 Let $t_e=null$ and $t_s=null$\\ 
            }
        }
    }
}
\For{each item $c_e$ that flagged as $e$ in $T_c$} 
{
    \If{there are $m$ continuous $s$ items on the left of $c$}
    {
        Let $c_s=$ the closest left $s$ item\\
        Put $(c_s,c_e)$ into $T$\\
    }
}
\end{algorithm}
\section {Experimental Evaluation}
\subsection {Experiment Setup}
We evaluate Dolha snapshot and Dolha persistent structure separately. 

In Dolha snapshot experiment, we compare Dolha snapshot with adjacency matrix in hash table and adjacency list in hash table. Since TCM is based on adjacency matrix in hash table and the java project GraphStream is based on adjacency list in hash table, we believe the comparison to these two general GraphStream structures could reflect the performance of Dolha properly. For the three structures, we first compare the average operation time cost and space cost and then compare the speed of query primitives.

We use the same hash function (MurmurHash) for all the structures and build the same vertex hash table and vertex table for all three structures so they all share the same vertex operation time cost and accuracy. Because the full adjacency matrix is too large, we compress the matrix in certain ratios that costs similar space as Dolha. That makes TCM become an approximation structure and we take account of the relative error. 

In Dolha persistent experiment, since there is no similar system for comparison, we build an adjacency list in hash table with an extra time line which stores all the edge update information. We use the adjacency list as baseline method to compare with Dolha persistent on the speed of sliding window update, query primitives and time related queries.

\subsubsection {Dataset}
\begin{enumerate}
	\item \textbf{DBLP \cite{dblp}:} DBLP dataset contains $1,482,029$ unique authors and $10,615,809$ time-stamped coauthorship edges between authors (about $6$ million unique edges). It’s a directed graph and we assign each streaming edge with weight $1$.
	\item \textbf{GTGraph \cite{gtgraph}:} We use the graph generator toll GTGraph to generate a directed graph. We use the R-MAT model generate a large network with power-law degree distributions and add weight $1$ to for each edge and use the system clock to get the time-stamp. The generated graph contains 30 million vertices and $1$ billion stream edges.
	\item \textbf{Twitter \cite{twitterdata}:} We use the Twitter link structure data with $56$ million vertices and $2$ billion edges as a directed streaming graph and assign weight 1 to each edge.
	\item \textbf{CAIDA \cite{caida}:}  CAIDA Internet Anonymized Traces 2015 Data-set obtained from www.caida.org. The network data contains $445,440,480$ communication records (edges) (about $100$ million unique edges) concerning $2,601,005$ different IP addresses (vertices).
\end{enumerate}
We use 4 datasets for Dolha snapshot experiment: The DBLP, GTGraph and Twitter are used for Dolha snapshot experiments and DBLP and CAIDA are used for Dolha persistent experiments.

\subsubsection {Environment}
All experiments are performed on a server with dual 8-core CPUs (Intel Xeon CPU E5-2640 v3 @ 2.60GHz) and 128 GB DRAM memory, running CentOS. All the data structures are implemented in C++.

\subsection {Dolha Snapshot Experimental Results}
\subsubsection {Construction}
Firstly, we compare the average processing time cost of stream graph on three structures and the space cost of them. In real world scenario, the insertion, deletion and update operations are usually coming randomly and the average stream processing speed is the key performance indicator of the system and all three operations time costs on Dolha are $O(1)$. So we load the datasets $2$ times as insertion and update and set the weight to $-3$ for last loading as deletion. Then we calculate the average time as the stream processing time cost and present it in the form of operations per second. During the data loading, we record the actual memory consuming when the edges are fully loaded. The results are showing in Figure \ref{fig:dj}.

\begin{figure}[h!]
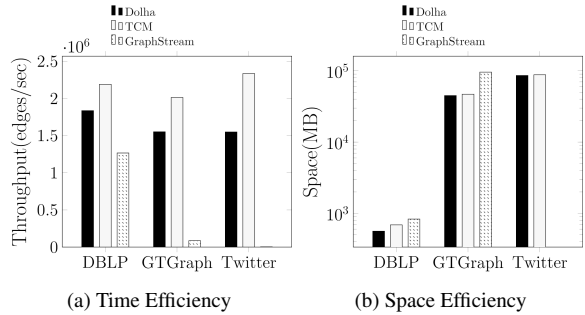

\centering
    \begin{subfigure}[t]{0.45\linewidth}
	    \centering
	    \resizebox{\linewidth}{!}
	    {
	        \includegraphics{time}
	    }
	    \caption{Time Efficiency}
	    \label{fig:time}
    \end{subfigure}
    \begin{subfigure}[t]{0.45\linewidth}
        \centering
        \resizebox{\linewidth}{!}
        {
            \includegraphics{space}
        }
        \caption{Space Efficiency}
        \label{fig:space}
    \end{subfigure}
\caption{Time and space cost for 3 streaming graph structure}
\label{fig:dj}
\end{figure}

In DBLP dataset, Dolha processing speed reaches $1,837,357$ operations per second which almost same as TCM’s $2,192,715$ operations per second and faster than GraphStream’s $1,266,815$ operations per second. Since the preset compress ratio, the memory cost of TCM is $690$MB which is similar to Dolha’s $563$MB. The GraphStream costs $833$MB which is worse than Dolha.
In GTGraph dataset, the performance remains the same. The TCM is the fastest structure with $2,014,768$ operations per second and Dolha is not far behind with $1,552,536$ operations per second. The speed of GraphStream drops significantly to $85,441$ operations per second and the space cost reaches $96$GB which is way higher than Dolha’s $45$GB and TCM’s $47$GB.
In Twitter dataset, the GraphStream runs out memory since the enormous space cost of sorted list maintenance. The performances of Dolha and TCM are steady. Dolha costs $86$GB memory and reaches $1,550,197$ operations per second while the TCM costs $88$GB and reaches $2,336,785$ operations per second.
The time cost results show that Dolha is slightly slower on stream processing speed than the TCM but significantly faster than the GraphStream. Since the TCM is an approximation structure and Dolha is an exact structure, the latency is acceptable. The space cost results show that Dolha could process 2 billion edges stream on less than $90$GB memory.

\subsubsection {Query Primitives}
In this part, we compare the query primitives speed on the three systems. The vertex query, the edge query, 1-hop successor query and 1-hop precursor query are taken into account. The time-related query and sliding window update are not supported by the other two structures and the time costs are depended on the given parameters, so we have not run experiment on these two queries. 

\textbf{Vertex Query}: The three structures share the same vertex hash table and vertex table, so the vertex query speeds are same. We run $25$ random vertex queries which cost $14,146$ nanoseconds in total. It means the average vertex query is $566$ nanoseconds per query.

\textbf{Edge Query}: We run $50$ random edge queries for three structures on each dataset. The results show that speed of edge query on Dolha is similar as on TCM with 0 relative error and much faster than GraphStream.

\begin{figure}[h!]
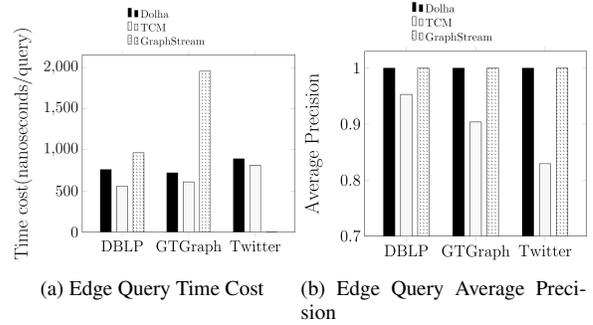

\centering
    \begin{subfigure}[t]{0.45\linewidth}
	    \centering
	    \resizebox{\linewidth}{!}
	    {
	        \includegraphics{edgequerytime}
	    }
	    \caption{Edge Query Time Cost}
	    \label{fig:dj:edgetime}
    \end{subfigure}
    \begin{subfigure}[t]{0.45\linewidth}
        \centering
        \resizebox{\linewidth}{!}
        {
            \includegraphics{edgequeryper}
        }
        \caption{Edge Query Average Precision }
        \label{fig:dj:edgeper}
    \end{subfigure}
\caption{Time cost and average precision for edge query}
\label{fig:dj}
\end{figure}

\textbf{1-hop Successor Query and 1-hop Precursor Query}:  We randomly choose $25$ vertices and run 1-hop successor query and 1-hop precursor query for three structures on each dataset. Since the query speed depends on the size of results set, we calculate the average query speed as nanoseconds per result. The TCM has almost $0$ average precision on these queries and slowest query speed. Among the threes structures, Dolha has the best performance with fast query speed and 100\% precision.

\begin{figure}[h!]
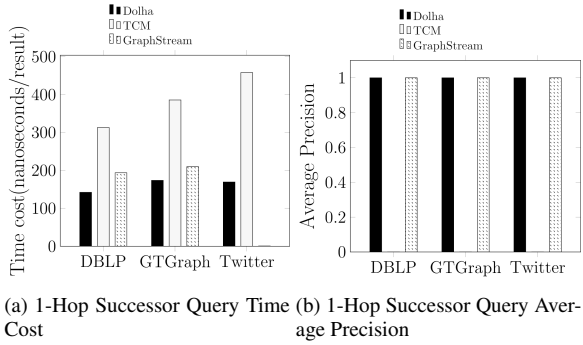

\centering
    \begin{subfigure}[t]{0.45\linewidth}
	    \centering
	    \resizebox{\linewidth}{!}
	    {
	        \includegraphics{successortime}
	    }
	    \caption{1-Hop Successor Query Time Cost}
	    \label{fig:dj:successortime}
    \end{subfigure}
    \begin{subfigure}[t]{0.45\linewidth}
        \centering
        \resizebox{\linewidth}{!}
        {
            \includegraphics{successorper}
        }
        \caption{1-Hop Successor Query Average Precision }
        \label{fig:dj:successorper}
    \end{subfigure}
\caption{Time cost and average precision for 1-hop successor query}
\label{fig:dj}
\end{figure}

\begin{figure}[h!]
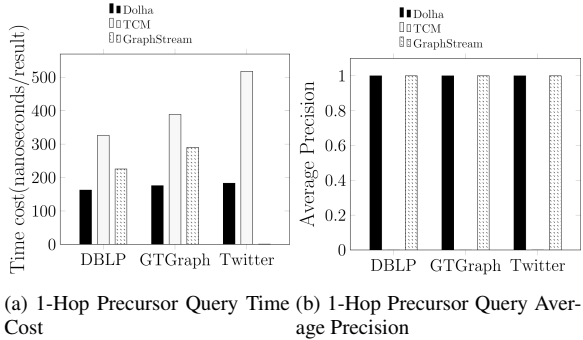

\centering
    \begin{subfigure}[t]{0.45\linewidth}
	    \centering
	    \resizebox{\linewidth}{!}
	    {
	        \includegraphics{precursortime}
	    }
	    \caption{1-Hop Precursor Query Time Cost}
	    \label{fig:dj:precursortime}
    \end{subfigure}
    \begin{subfigure}[t]{0.45\linewidth}
        \centering
        \resizebox{\linewidth}{!}
        {
            \includegraphics{precursorper}
        }
        \caption{1-Hop Precursor Query Average Precision }
        \label{fig:dj:precursorper}
    \end{subfigure}
\caption{Time cost and average precision for 1-hop precursor query}
\label{fig:dj}
\end{figure}

Compare to the GraphStream, Dolha has great advantages on the average stream processing time cost, space cost, edge query speed, 1-hop successor query and 1-hop precursor query speed. Dolha is slightly slower than the TCM with similar space cost on average stream processing time cost, space cost, edge query speed but faster on 1-hop successor query and 1-hop precursor query. On the other hand, the Dolha is an exact structure and the TCM is an approximation structure. 

\textbf{Directed Triangle Finding:}
We run continuous directed triangle finding algorithm on DBLP and GTGraph 1 billion date set using Dolha snapshot and GraphStream. For DBLP dataset, Dolha could process 759,866 edge updates per-second and GraphStream only could process 238,095 edge updates per-second. For GTGraph 1 billion date set, Dolha could deal 129,853 throughput edges per-second but GraphStream could only deal less than 10,000 throughput edges per-second.

\subsection {Dolha Persistent Experimental Results}

\subsubsection {Construction and Sliding Window Update}
We set window length $=\frac{1}{10}|S|$, slide length $=\frac{1}{5}$ window length as W1 and slide length $=\frac{1}{50}$ window length as W2. Then we load the DBLP and CAIDA dataset with and without sliding window update. Figure \ref{fig:timewindow} shows the through-puts of Dolha persistent and adjacency list plus time-line with and without sliding window update. 

On DBLP date set, Dolha persistent reaches $2,008,420$ edges update per second without sliding window update, $1,979,889$ edges update per second in W1 and $1,961,238$ edges update per second in W2. The adjacency list plus time-line only can process $1,120,269$ edges update per second without sliding window update, $893,795$ edges update per second in W1 and $583,367$ edges update per second in W2.

On CAIDA dataset, Dolha persistent reaches $3,969,514$ edges update per second without sliding window update, $3,917,037$ edges update per second in W1 and $3,425,009$ edges update per second in W2. The results are way better than the adjacency list plus time-line's speeds: $761,834$ edges update per second without sliding window update, $676,077$ edges update per second in W1 and $472,953$ edges update per second in W2. 

The construction time costs in different window setting on Dolha persistent are similar which means the the size of slide length are insignificant to the edge processing. The outstanding high speed is caused by the high duplicated edge rate on CAIDA dataset. We set the edge hash table same size as edge table, but the unique edge number is only $\frac{1}{4}$ of total stream edge number which reduces the hash collision significantly. And when we process the duplicate edge update, we do not need to check the vertices by using vertex hash table.

 \begin{figure}[h!]
\centering
\resizebox{0.8\linewidth}{!}{
	\includegraphics{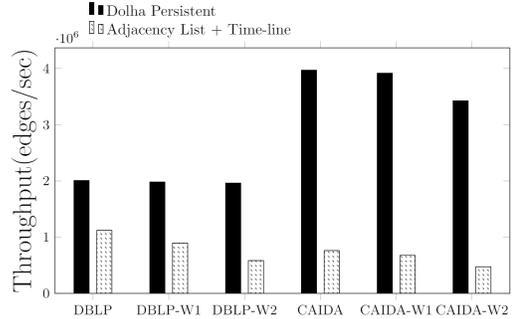}
}
\caption{Edge throughput without and with time window update}
\label{fig:timewindow}
\end{figure}
\subsubsection {Query Primitives}
The query primitives of DBLP on Dolha persistent are exact the same as Dolha snapshot, we only compare the CAIDA with adjacency list plus time-line.

\textbf{Vertex Query: }The two structures use the same vertex hash table and vertex table. We run $25$ random vertex queries and the average vertex query is $605$ nanoseconds per query.

\textbf{Edge Query: }We run $50$ random edge queries on both data structures. The result shows that Dolha persistent is $5$ times faster than adjacency list plus time-line.

\textbf{1-hop Successor Query and 1-hop Precursor Query: }We randomly choose $25$ vertices and run 1-hop successor query and 1-hop precursor query on two structures. Dolha persistent is slighly faster than adjacency list plus time-line.

 \begin{figure}[h!]
\centering
\resizebox{0.7\linewidth}{!}{
	\includegraphics{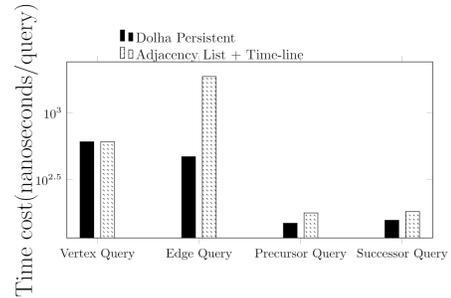}
}
\caption{Query primitives on CAIDA}
\label{fig:persistentquery}
\end{figure}

\subsubsection {Time Related Queries}
\begin{figure}[h!]
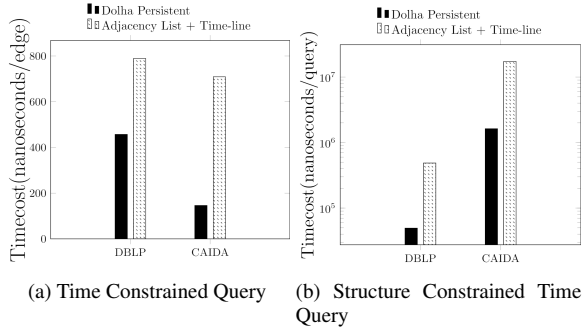

\centering
    \begin{subfigure}[t]{0.45\linewidth}
        \centering
        \resizebox{\linewidth}{!}
        {
            \includegraphics{patterne}
        }
        \caption{Time Constrained Query}
        \label{fig:dj:pattern}
    \end{subfigure}
    \begin{subfigure}[t]{0.45\linewidth}
	    \centering
	    \resizebox{\linewidth}{!}
	    {
	        \includegraphics{timequerye}
	    }
	    \caption{Structure Constrained Time Query}
	    \label{fig:dj:timequery}
    \end{subfigure}

\caption{Time related query}
\label{fig:dj}
\end{figure}

\textbf{Time Constrained Pattern Query:}
For time constrained pattern query, we randomly choose $3$ pairs of time-stamps as time constrain and extract the eligible edges to form a candidate subgraph. Figure \ref{fig:dj:pattern} shows the average candidate subgraph forming speeds of Dolha persistent and adjacency list plus time-line. In DBLP, we reach $457$ nanoseconds per edge to extract the candidate subgraph into a Dolha snapshot and the adjacency list plus time-line can only construct $789$ nanoseconds per edge into an adjacency list. In CAIDA, the speed reaches $146$ nanoseconds per edge and the adjacency list plus time-line can only process $709$ nanoseconds per edge. 

\textbf{Structure Constrained Time Query:}
To compare structure constrained time query, we randomly choose $5$ query edge sets and each set has $5$ edges. The average query time of Dolha persistent is $49,378$ nanoseconds per query on DBLP and $1,623,200$ nanoseconds per query on CAIDA. The average query time of adjacency list plus time-line is $486,576$ nanoseconds per query on DBLP and $17,312,871$ nanoseconds per query on CAIDA. 

\section {Conclusions}
\label{sec:con}
We have proposed an exact streaming graph structure Dolha which could maintain high speed and high volume streaming graph in linear time cost and near linear space cost. We have shown that Dolha is a general propose structure that could support the query primitives which are the cornerstone of common graph algorithms. We also present the Dolha persistent structure which could support sliding window update and time related queries. The experiment results have proved that Dolha has better performance than the other streaming graph structures.

\clearpage
\balance

\bibliographystyle{IEEEtran}
\bibliography{main}  

\begin{thebibliography}{10}
\providecommand{\url}[1]{#1}
\csname url@samestyle\endcsname
\providecommand{\newblock}{\relax}
\providecommand{\bibinfo}[2]{#2}
\providecommand{\BIBentrySTDinterwordspacing}{\spaceskip=0pt\relax}
\providecommand{\BIBentryALTinterwordstretchfactor}{4}
\providecommand{\BIBentryALTinterwordspacing}{\spaceskip=\fontdimen2\font plus
\BIBentryALTinterwordstretchfactor\fontdimen3\font minus
  \fontdimen4\font\relax}
\providecommand{\BIBforeignlanguage}[2]{{%
\expandafter\ifx\csname l@#1\endcsname\relax
\typeout{** WARNING: IEEEtran.bst: No hyphenation pattern has been}%
\typeout{** loaded for the language `#1'. Using the pattern for}%
\typeout{** the default language instead.}%
\else
\language=\csname l@#1\endcsname
\fi
#2}}
\providecommand{\BIBdecl}{\relax}
\BIBdecl

\bibitem{graphsurvey2012vldb}
S.~Guha and M.~Andrew, ``Graph synopses, sketches, and streams: a survey,''
  \emph{PVLDB}, vol.~5, no.~12, pp. 2030--2031, 2012.

\bibitem{twitter}
``Tweet statistics,''
  \url{http://expandedramblings.com/index.php/march-2013-by-the-numbers-a-few-amazingtwitter-stats/10/}.

\bibitem{email}
``Email statistics report, 2015-2019,''
  \url{https://radicati.com/wp/wp-content/uploads/2015/02/Email-Statistics-Report-2015-2019-Executive-Summary.pdf}.

\bibitem{tcm}
N.~Tang, Q.~Chen, and P.~Mitra, ``Graph stream summarization: From big bang to
  big crunch,'' \emph{SIGMOD}, pp. 1481--1496, 2016.

\bibitem{triest}
L.~De~Stefani, A.~Epasto, M.~Riondato, and E.~Upfal, ``{TRI\`{E}ST}: Counting
  local and global triangles in fully-dynamic streams with fixed memory size,''
  in \emph{Proceedings of the 22nd ACM SIGKDD International Conference on
  Knowledge Discovery and Data Mining}, ser. KDD '16.\hskip 1em plus 0.5em
  minus 0.4em\relax ACM, 2016.

\bibitem{timingsubg}
Y.~Li, L.~Zou, M.~T. Ozsu, and D.~Zhao, ``Time constrained continuous subgraph
  search over streaming graphs,'' \url{https://arxiv.org/pdf/1801.09240.pdf},
  2018.

\bibitem{cycledetection2018vldb}
X.~Qiu, W.~Cen, Z.~Qian, Y.~Peng, Y.~Zhang, X.~Lin, and J.~Zhou, ``Real-time
  constrained cycle detection in large dynamic graphs,'' \emph{Proceedings of
  the VLDB Endowment}, vol.~11, no.~12, 2018.

\bibitem{graphstreamproject}
Y.~Pigne, A.~Dutot, F.~Guinand, and D.~Olivier, ``Graphstream: A tool for
  bridging the gap between complex systems and dynamic graphs,'' \emph{EPNACS},
  2007.

\bibitem{gmatrix}
A.~Khan and C.~C. Aggarwal, ``Query-friendly compression of graph streams,''
  \emph{IEEE/ACM International Conference on Advances in Social Networks
  Analysis and Mining}, pp. 130--137, 2016.

\bibitem{hyperanf}
P.~Boldi, M.~Rosa, and S.~Vigna, ``Hyperanf: approximating the neighbourhood
  function of very large graphs on a budget,'' \emph{International world wide
  web conferences}, pp. 625--634, 2011.

\bibitem{gao2014continuous}
J.~Gao, C.~Zhou, J.~Zhou, and J.~X. Yu, ``Continuous pattern detection over
  billion-edge graph using distributed framework,'' in \emph{Proc. 30th IEEE
  International Conference on Data Engineering}, 2014, pp. 556--567.

\bibitem{bloom}
A.~Z. Broder and M.~Mitzenmacher, ``Network applications of bloom filters: A
  survey,'' \emph{Internet Mathematics}, vol.~1, no.~4, pp. 485--509, 2004.

\bibitem{countmin}
G.~Cormode and S.~Muthukrishnan, ``An improved data stream summary: The
  count-min sketch and its applications,'' \emph{latin american symposium on
  theoretical informatics.}, pp. 29--38, 2004.

\bibitem{streamsurvey}
A.~Mcgregor, ``Graph stream algorithms: a survey,'' \emph{SIGMOD Record},
  vol.~43, no.~1, pp. 9--20, 2014.

\bibitem{slidingwindow}
L.~Gao, L.~Golab, M.~T. Ozsu, and G.~Aluc, ``Stream watdiv: A streaming rdf
  benchmark,'' no.~3, 2018.

\bibitem{hashcollision}
C.~Stein, S.~Drysdale, and K.~Borgart, ``Probability calculations in hashing,''
  in \emph{Discrete Mathematics for Computer Scientists}.\hskip 1em plus 0.5em
  minus 0.4em\relax Addison-Wesley; 1st edition, 2010, pp. 245--254.

\bibitem{triangle}
T.~Schank and D.~Wagner, ``Finding, counting and listing all triangles in large
  graphs, an experimental study,'' in \emph{Nikoletseas S.E. (eds) Experimental
  and Efficient Algorithms. Lecture Notes in Computer Science}, vol. 3503,
  2005.

\bibitem{dblp}
E.~Demaine and M.~Hajiaghayi, ``Bigdnd: Big dynamic network data,''
  \url{http://projects.csail.mit.edu/dnd/DBLP/}.

\bibitem{gtgraph}
``Gtgraph: A suite of synthetic random graph generators,''
  \url{http://www.cse.psu.edu/~kxm85/software/GTgraph/}.

\bibitem{twitterdata}
M.~Cha, H.~Haddadi, F.~Benevenuto, and K.~P. Gummadi, ``{Measuring User
  Influence in Twitter: The Million Follower Fallacy}.''

\bibitem{caida}
``Caida internet anonymized traces 2015 dataset,''
  \url{http://www.caida.org/home/}.

\end{thebibliography}

\end{document}